\documentclass[aps,prr,twocolumn,superscriptaddress]{revtex4}
\usepackage{graphicx}
\usepackage{dcolumn}
\usepackage{bm}
\usepackage{physics}
\usepackage{amsmath}
\usepackage{amssymb}
\usepackage{array}
\usepackage{color}
\newcolumntype{P}[1]{>{\centering\arraybackslash}p{#1}}
\usepackage[colorlinks=true, breaklinks=true, linkcolor=blue, citecolor=blue, urlcolor=blue]{hyperref}

\newcommand{\geneva}{Department of Quantum Matter Physics, University of Geneva, Quai Ernest-Ansermet 24, 1211 Geneva, Switzerland}

\usepackage{ifthen}
\usepackage{booktabs}
\usepackage{multirow}

\begin{document}

\title{Bose-Hubbard triangular ladder in an artificial gauge field}

\date{\today}

\begin{abstract}
We consider interacting bosonic particles on a two-leg triangular ladder in the presence of an artificial gauge field. We employ density matrix renormalization group numerical simulations and analytical bosonization calculations to study the rich phase diagram of this system. We show that the interplay between the frustration induced by the triangular lattice geometry and the interactions gives rise to multiple chiral quantum phases. Phase transition between superfluid to Mott-insulating states occur, which can have Meissner or vortex character. Furthermore, a state that explicitly breaks the symmetry between the two legs of the ladder, the biased chiral superfluid, is found for values of the flux close to $\pi$. In the regime of hardcore bosons, we show that the extension of the bond order insulator beyond the case of the fully frustrated ladder exhibits Meissner-type chiral currents. We discuss the consequences of our findings for experiments in cold atomic systems. 
\end{abstract}
\author{Catalin-Mihai Halati}
\affiliation{\geneva}
\author{Thierry Giamarchi}
\affiliation{\geneva}
\maketitle

\section{Introduction \label{sec:intro}}

The interplay between kinetic energy and interactions leads, for quantum systems, to a very rich set of many-body phases with remarkable properties, such as superconductivity, or Mott insulators. This is particularly true in reduced dimensionality, where the effects of interactions are at their maximum. This leads in one dimension to a set of properties, known as Tomonaga-Luttinger liquids \cite{Giamarchibook}. These are quite different from the typical physics that exists in higher dimensions, characterized by ordered states with single particle type excitations, such as Bogoliubov excitations for bosons, or Landau quasiparticles for fermions. 

An intermediate situation is provided by ladders, i.e. a small number of one-dimensional (1D) chains coupled by tunneling. Such systems possess some unique properties, different from both the one-and the high-dimensional ones. For example fermionic ladders exhibit superconductivity with purely repulsive interactions, at variance with isolated 1D chains that are dominated by antiferromagnetic correlations \cite{dagotto_1996_ladder_review}. 

Ladders are also the minimal systems in which the orbital effects of a magnetic field can be explored. For bosonic ladders this has allowed to predict \cite{OrignacGiamarchi2001} the existence of quantum phase transitions as a function of the flux between a low field phase with current along the legs (Meissner phase) and a high field phase with currents across the rungs and the presence of vortices (vortex phase), akin to the transition occurring in type II superconductors. 
Ultracold atomic systems offer the possibility of studying such systems coupled to artificial gauge fields \cite{DalibardOehberg2011, GoldmanSpielman2014}, and the Meissner to vortex phase transition has been observed experimentally \cite{AtalaBloch2014}.
These works have paved the way for a flurry of studies for other situations both for bosonic and fermionic ladders \cite{OrignacGiamarchi2001,CarrNersesyan2006, RouxPoilblanc2007, DharParamekanti2012, DharParamekanti2013, PetrescuLeHur2013, WeiMueller2014, TokunoGeorges2014, DiDioChiofalo2015, PiraudSchollwoeck2015, GreschnerVekua2015, PetrescuLeHur2015, UchinoTokuno2015, GreschnerVekua2016, Uchino2016, OrignacChiofalo2016, PetrescuLeHur2017, StrinatiMazza2017, OrignacDePalo2017, Strinati_2019_fermionicladder, Buser_2019_ladderfiniteT, Buser_2020_ladderquench, Qiao_2021_biasedladder, Haller_2018_fermionic_flux_ladder, Haller_2020_bosonic_flux_ladder}. Furthermore, properties beyond the phase diagram, such as the Hall effect, were also studied \cite{Greschner_2019_HallEffect,Buser_2021_HallEffect} and even measured \cite{ManciniFallani2015, Genkina_2019_Hofstadterribbons, Zhou_2022_Hallexperiment}. 

These extensive studies of ladders have however concentrated mostly on square ladders, for which the effect of hopping is unfrustrated, leaving the case of triangular ladders under flux relatively unexplored, despite some previous studies focusing on particular setups, or corners of the phase diagram \cite{MishraParamekanti2013, MishraMukerjee2014,AnisimovasJuzeliunas2016,AnGadway2018,AnGadway2018b, Romen_2018_chiralmottfrustration,GreschnerMishra2019, CabedoCeli2020,LiLi2020, Roy_2022_frustrated1dbosons}. The triangular structure is not bipartite and, thus, prevents the particle-hole symmetry that occurs naturally in square lattices. This has drastic consequences since it leads to frustration of the kinetic energy and, thus, to quite different properties, as was largely explored for two-dimensional systems \cite{Wessel_2005_supersolidtriangular, Becker_2010_triangularlattice,StruckSengstock2011, EckardtLewenstein2011,Zaletel_motttriangular}. 

In this paper, we explore the phase diagram of a triangular two leg bosonic ladder under an artificial magnetic field. We consider bosons with a contact repulsive interaction. We study, using a combination of analytical bosonization and numerical density matrix renormalization group (DMRG) techniques the phase diagram of such a system as a function of the magnetic field, filling and repulsion between the bosons. We discuss in particular our findings in comparison with the phases found for the square ladders. 

The plan of the paper is as follows, in Sec.~\ref{sec:model} we describe the model considered, its non-interacting limit and the observables of interest. In Sec.~\ref{sec:methods} we briefly discuss the methods employed in this work. We present the results regarding the phase diagram at half filling in Sec.~\ref{sec:half_filling}. In this regime, we identify the following quantum phases, the Meissner superfluid (M-SF), the vortex superfluid (V-SF) and the biased chiral superfluid (BC-SF), which breaks the $\mathbb{Z}_2$ symmetry of the ladder. For the fully frustrated $\pi$-flux ladder, Sec.~\ref{sec:2chain_pi}, we obtain a transition between superfluid and chiral superfluid states.
In the limit of hardcore bosons, Sec.~\ref{sec:hardcore}, at $\pi$ flux we have successive phase transitions between superfluid, bond order insulator and chiral superfluid states.
The bond order extends in the phase diagram for lower values of the flux to the chiral bond order insulator (C-BOI).
At unity filling for interacting bosons, Sec.~\ref{sec:unity_filling}, also a Meissner Mott insulator (M-MI) can be found in the phase diagram.
We discuss our results in Sec.~\ref{sec:discussions} and conclude in Sec.~\ref{sec:conclusions}.

\begin{figure}[!hbtp]
\centering
\includegraphics[width=.48\textwidth]{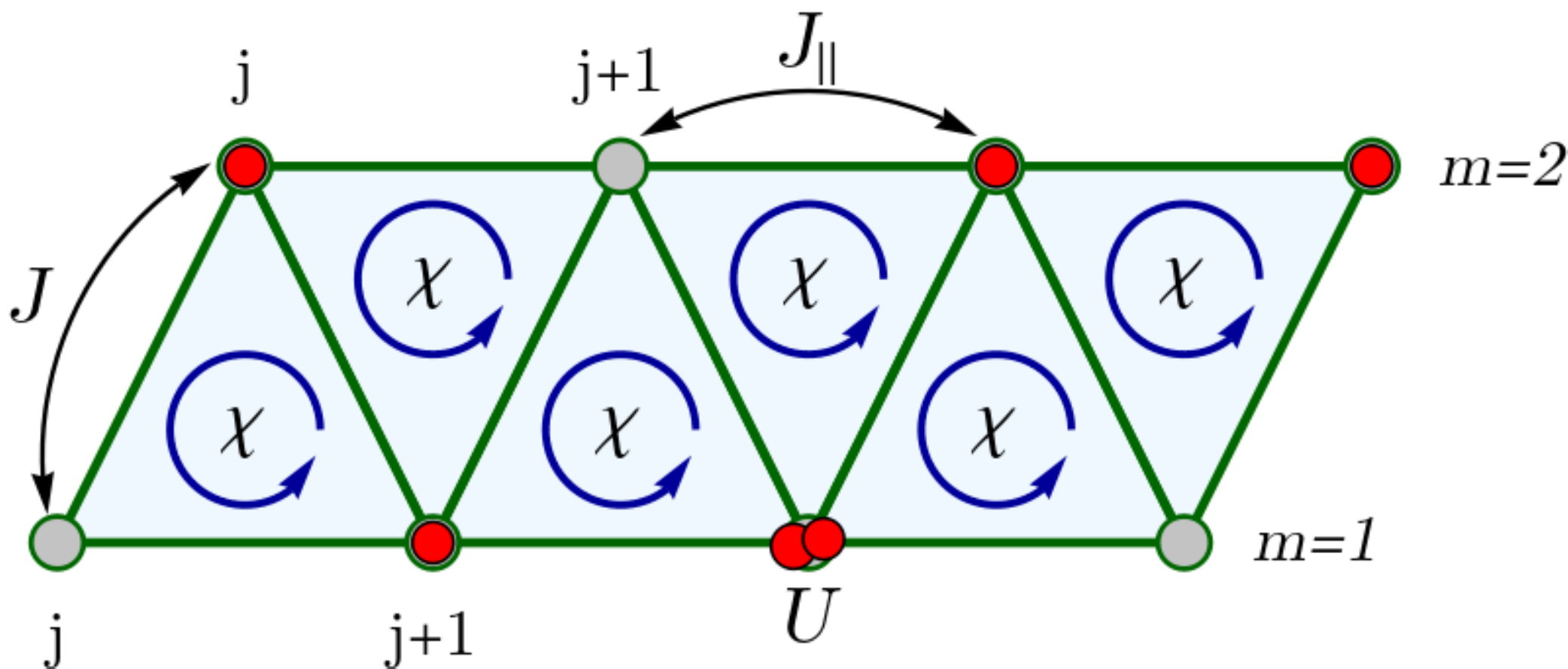}
\caption{Sketch of the setup. The bosonic atoms are confined in a quasi-one-dimensional triangular ladder. The legs are numbered by $m=1,2$ and the sites on each leg by $j$.
The atoms tunnel along the legs with the amplitude $J_\|$, along the rungs with the amplitude $J$ and have an on-site interaction of strength $U$. Each triangular plaquette is pierced by a flux $\chi$.
 }
\label{fig:model}
\end{figure}

\section{Model \label{sec:model}}

\subsection{Setup \label{sec:setup}}

We consider interacting bosonic atoms confined to a triangular ladder in an artificial gauge field, as sketched in Fig.~\ref{fig:model}. The Bose-Hubbard Hamiltonian of the system is given by
\begin{align} 
\label{eq:Hamiltonian}
 H=& H_\parallel+H_\perp+H_{\text{int}},\\
H_\parallel = & -J_\| \sum_{j=1}^{L-1} \left( e^{-i\chi}b^\dagger_{j,1}b_{j+1,1}+e^{i\chi} b^\dagger_{j,2}b_{j+1,2} + \text{H.c}. \right), \nonumber \\
H_\perp = & -J \sum_{j=1}^L \left( b^\dagger_{j,1}b_{j,2} + \text{H.c}. \right) \nonumber \\
&-J \sum_{j=1}^{L-1} \left(b^\dagger_{j+1,1}b_{j,2} + \text{H.c}. \right),\nonumber \\
H_{\text{int}}=&\frac{U}{2} \sum_{j=1}^L \sum_{m=1}^2 n_{j,m}(n_{j,m}-1). \nonumber
\end{align}
The bosonic operator $b_{j,m}$ and $b^\dagger_{j,m}$ are the annihilation and creation operators of the particles at position $j$ and leg $m=1,2$. We consider a total number of $N=\sum_{j=1}^L \sum_{m=1}^2 n_{j,m}$ atoms and that the ladder has $L$ sites on each leg. The atomic density is given by $\rho=N/(2L)$. $H_\|$ describes the tunneling along the two legs of the ladder, indexed by $j$, with amplitude $J_\|$. The complex factor in the hopping stems from the artificial magnetic field, with flux $\chi$ \cite{DalibardOehberg2011, GoldmanSpielman2014}.
The tunneling along the rungs of the ladder is given by $H_\perp$ and has amplitude $J$. The atoms interact repulsively with with an on-site interaction strength $U>0$.

\subsection{Non-interacting limit \label{sec:noninteracting}}

\begin{figure}[!hbtp]
\centering
\includegraphics[width=.48\textwidth]{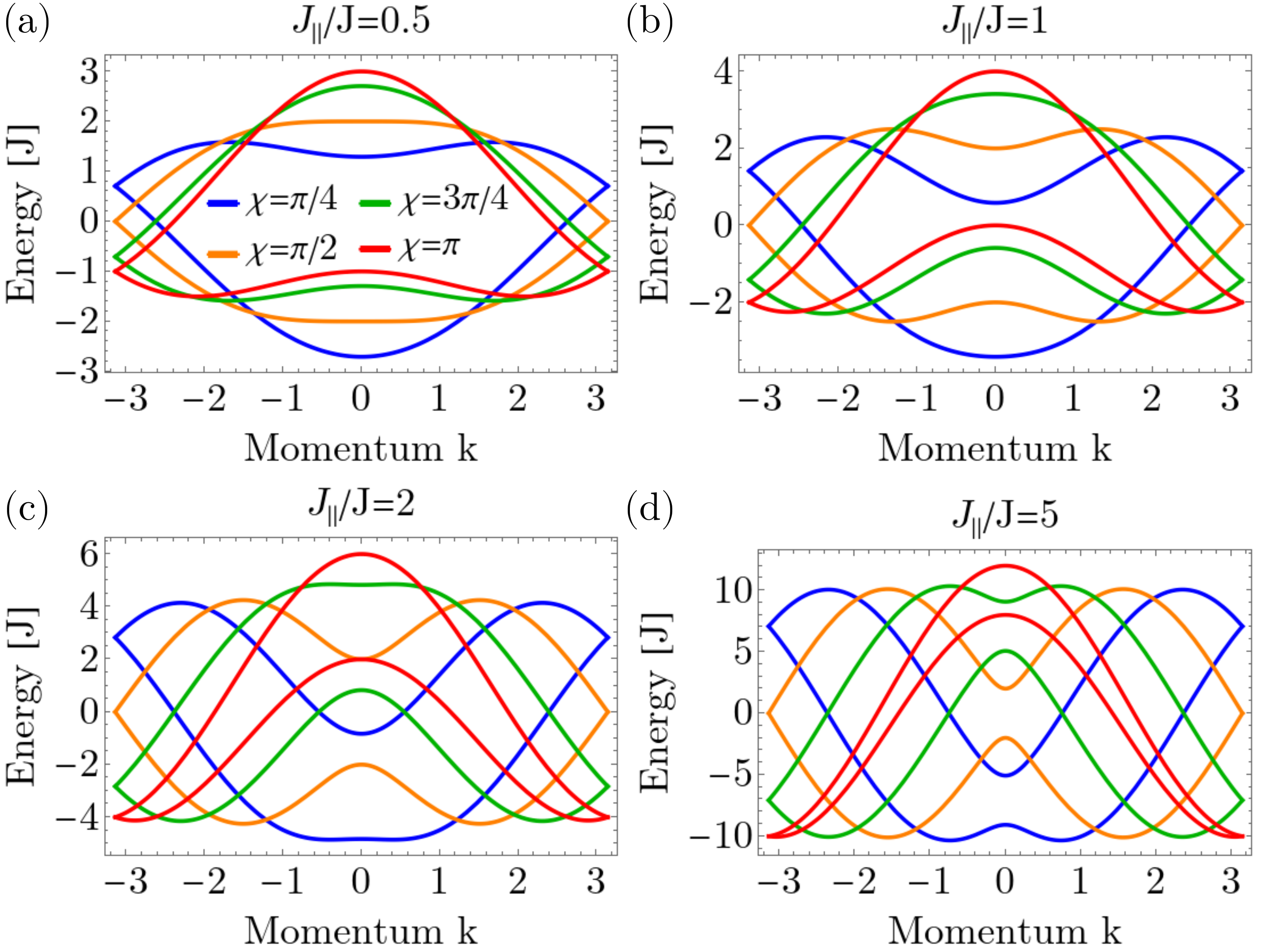}
\caption{\label{fig:bands}Single particle dispersion of model (\ref{eq:Hamiltonian}), $E_\pm(k)$ (\ref{eq:eigenvalues}), as a function of momentum $k$, for (a) $J_\|/J=0.2$, (b) $J_\|/J=1$, (c) $J_\|/J=2$, (d) $J_\|/J=5$, and different values of the flux $\chi\in\{\pi/4, \pi/2, 3\pi/4,\pi\}$. Note the presence of the two minima away from $k=0$ for some values of the flux and transverse hopping.}
\end{figure}

In the non-interacting, $U=0$, limit we can exactly diagonalize the Hamiltonian (\ref{eq:Hamiltonian}) (see Appendix~\ref{app:noninteracting}) and obtain the following dispersion relation
\begin{align} 
\label{eq:eigenvalues}
E_\pm (k)=&-2 J_\| \cos(k)\cos(\chi) \\
&\pm\sqrt{2J^2\left[1+\cos(k)\right]+4J_\|^2\sin^2(k)\sin^2(\chi)}. \nonumber
\end{align}
The non-interacting bands, $E_\pm (k)$, are represented in Fig.~\ref{fig:bands} for several values of $J_\|/J$ and $\chi$. We can observe that the lower band can have either a single minimum at $k=0$, e.g. for Fig.~\ref{fig:bands}(a) for $J_\|/J=0.5$ and $\chi=0.25 \pi$, or two minima at finite values of $k$, e.g. for Fig.~\ref{fig:bands}(c) for $J_\|/J=2$ and $\chi=0.5 \pi$. The position of the double minima depends on $J_\|/J$ and $\chi$.

The topology of the lower band can already provide some hints regarding the nature of the ground state in the case of weakly interacting bosons. Similarly with the analysis performed in the case of the square ladder with flux \cite{WeiMueller2014, UchinoTokuno2015} we expect phases of the following natures: Meissner states in the case in which the bosons condense in the $k=0$ minimum; vortex phases, in the case of two condensates in the two minima of the lower band; and states which break the $\mathbb{Z}_2$ symmetry of the ladder, corresponding to a condensate in just one of the double minima.
In Sec.~\ref{sec:half_filling} and Sec.~\ref{sec:unity_filling} we show how these states are realized on the triangular ladder in the interacting regime.

\subsection{Observables of interest \label{sec:observables}}

In the rest of this section, we describe some of the observables which are suitable for the investigation of the chiral phases we obtain in this system.
We define the local currents on the leg $j^\|_{j,m}$ and the rung $j^\perp_{j}$, respectively, as
\begin{align}
\label{eq:localcur}
&j^\|_{j,m} = -i J_\|\left( e^{i\chi (-1)^m} b_{j,m}^\dagger b_{j+1,m} -\text{H.c.} \right), \\
&j^\perp_{2j-1} = -i J (b_{j,1}^\dagger b_{j,2} -\text{H.c.}), \nonumber \\
&j^\perp_{2 j} = -i J (b_{j+1,1}^\dagger b_{j,2} -\text{H.c.}). \nonumber
\end{align}
In addition to the local currents, the chiral current $J_c$ and the average rung current $J_r$ are of interest and defined as
\begin{align}
\label{eq:cur}
&J_c = \frac{1}{2 (L-1)} \sum_j \left\langle j^\|_{j,1} - j^\|_{j,2} \right\rangle,\\ 
&J_r=\frac{1}{2 L-1} \sum_j \left|\left\langle j^\perp_{j}\right\rangle\right|. \nonumber
\end{align}

In order to identify biased phases, in which the $\mathbb{Z}_2$ symmetry between the two legs of the ladder is broken, we compute the density imbalance
\begin{align}
\label{eq:dn}
\Delta n =\frac{1}{2L}\sum_j \left(n_{j,1}-n_{j,2}\right).
\end{align}

Furthermore, we compute the central charge $c$, which can be interpreted as the number of gapless modes. We extract the central charge from the scaling of the von Neumann entanglement entropy $S_{vN}(l)$ of an embedded subsystem of length $l$ in a chain of length $L$. For open boundary conditions the entanglement entropy for the ground state of gapless phases is given by \cite{VidalKitaev2003,CalabreseCardy2004, HolzeyWilczek1994}
\begin{equation}
\label{eq:entropy}
S_{vN}=\frac{c}{6}\log\left(\frac{L}{\pi}\sin\frac{\pi l}{L}\right)+s_1,
\end{equation}
where $s_1$ is a non-universal constant, and we neglect  
logarithmic corrections \cite{AffleckLudwig1991} and oscillatory terms \cite{LaflorencieAffleck2006} due to the finite size of the system.

\section{Methods \label{sec:methods}}

\subsection{Bosonization \label{sec:bosonization}}

The low-energy physics of one dimensional interacting quantum systems, corresponding to the Tomonaga-Luttinger liquids universality class, can be described in terms of two bosonic fields $\phi$ and $\theta$ \cite{Giamarchibook}. These bosonic fields are related to the collective excitations of density and currents and fulfill the canonical commutation relation, $\left[ \phi(x),\nabla\theta(x') \right]=i\pi\delta(x-x')$.
In the bosonized representation, the single particle operator of the bosonic atoms can be written as \cite{Giamarchibook}:
\begin{align} 
\label{eq:bosonization}
b^\dagger_j&=(a\rho)^\frac{1}{2} \left\{ e^{-i \theta (a j)}+\sum_{p\neq 0} e^{i 2p \left[ \pi\rho aj -\phi(aj)\right]}e^{-i\theta(aj)}\right\},
\end{align}
with $\rho$ the density and $a$ the lattice spacing. In the following, we take $a=1$.

\subsection{MPS ground state simulations \label{sec:mps_gs}}

The numerical results were obtained using a finite-size density matrix renormalization group (DMRG) algorithm in the matrix product state (MPS) representation \cite{White1992, Schollwoeck2005, Schollwoeck2011, Hallberg2006, Jeckelmann2002}, implemented using the ITensor Library \cite{FishmanStoudenmire2020}. 
We compute the ground state of the model (\ref{eq:Hamiltonian}) for ladders with a number of rungs between $L=60$ and $L=180$, and with a maximal bond dimension up to 1800. This ensures that the truncation error is at most $10^{-9}$. 
Since we are considering a bosonic model with finite interactions the local Hilbert space is very large, thus, a cutoff for its dimension is needed. We use a maximal local dimension of at least four or five bosons per site. 
We checked that the local states with a higher number of bosons per site do not have an occupation larger than $10^{-5}$ for the parameters considered.
We make use of good quantum numbers in our implementation as the number of atoms is conserved in the considered model.

\section{Phase diagram at half filling, $\rho=0.5$ \label{sec:half_filling}}

\begin{figure}[!hbtp]
\centering
\includegraphics[width=.48\textwidth]{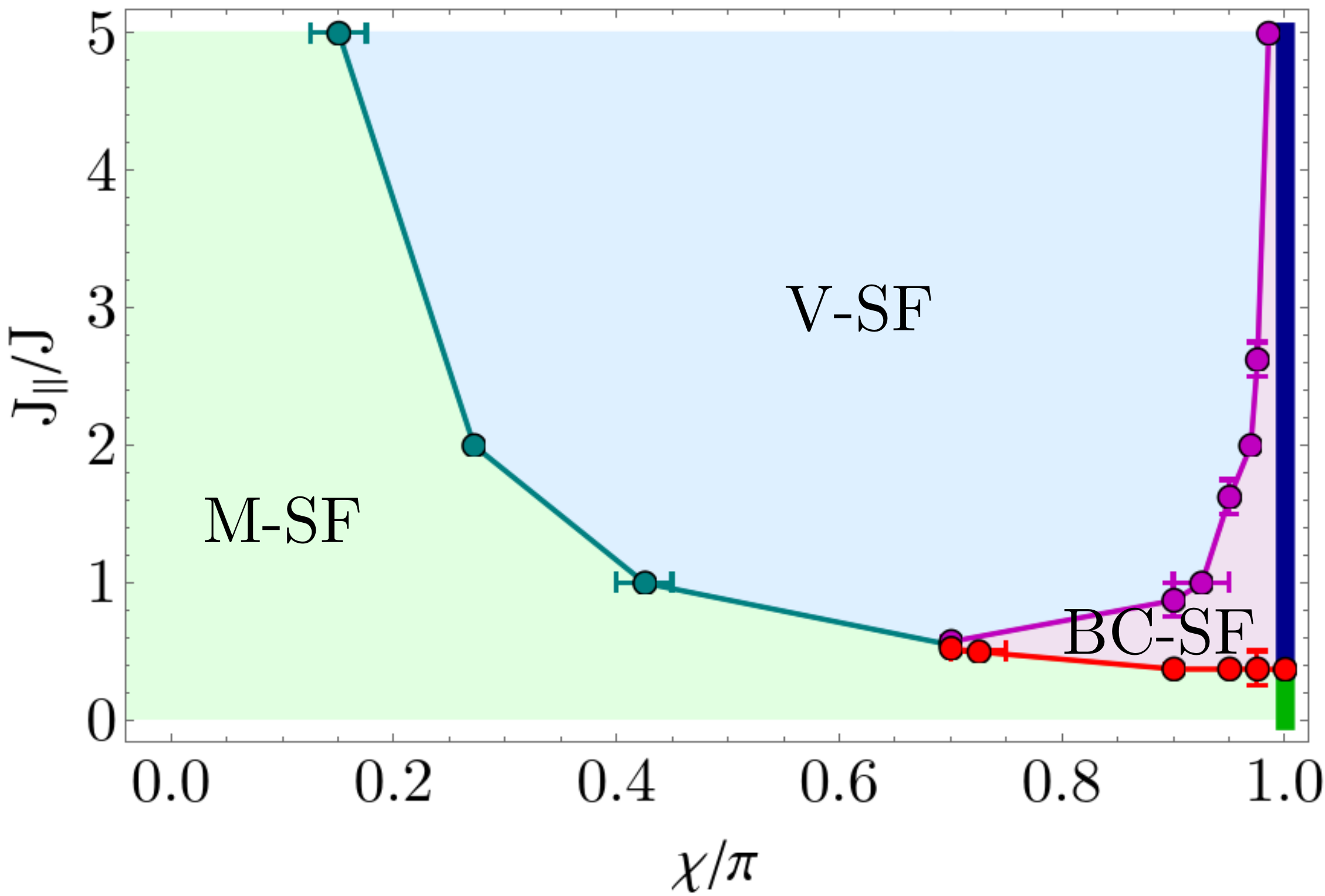}
\caption{Sketch of the phase diagram for half-filling, $\rho=0.5$, for $U/J=2.5$.
The identified phases (see text) are the Meissner superfluid (M-SF), the vortex superfluid (V-SF) and the biased chiral superfluid (BC-SF).
At $\chi=\pi$ (marked by a thick vertical line) we have a phase transition between a superfluid (green) and a chiral superfluid (dark blue).
 }
\label{fig:phasediag_n05}
\end{figure}

\begin{figure}[!hbtp]
\centering
\includegraphics[width=.48\textwidth]{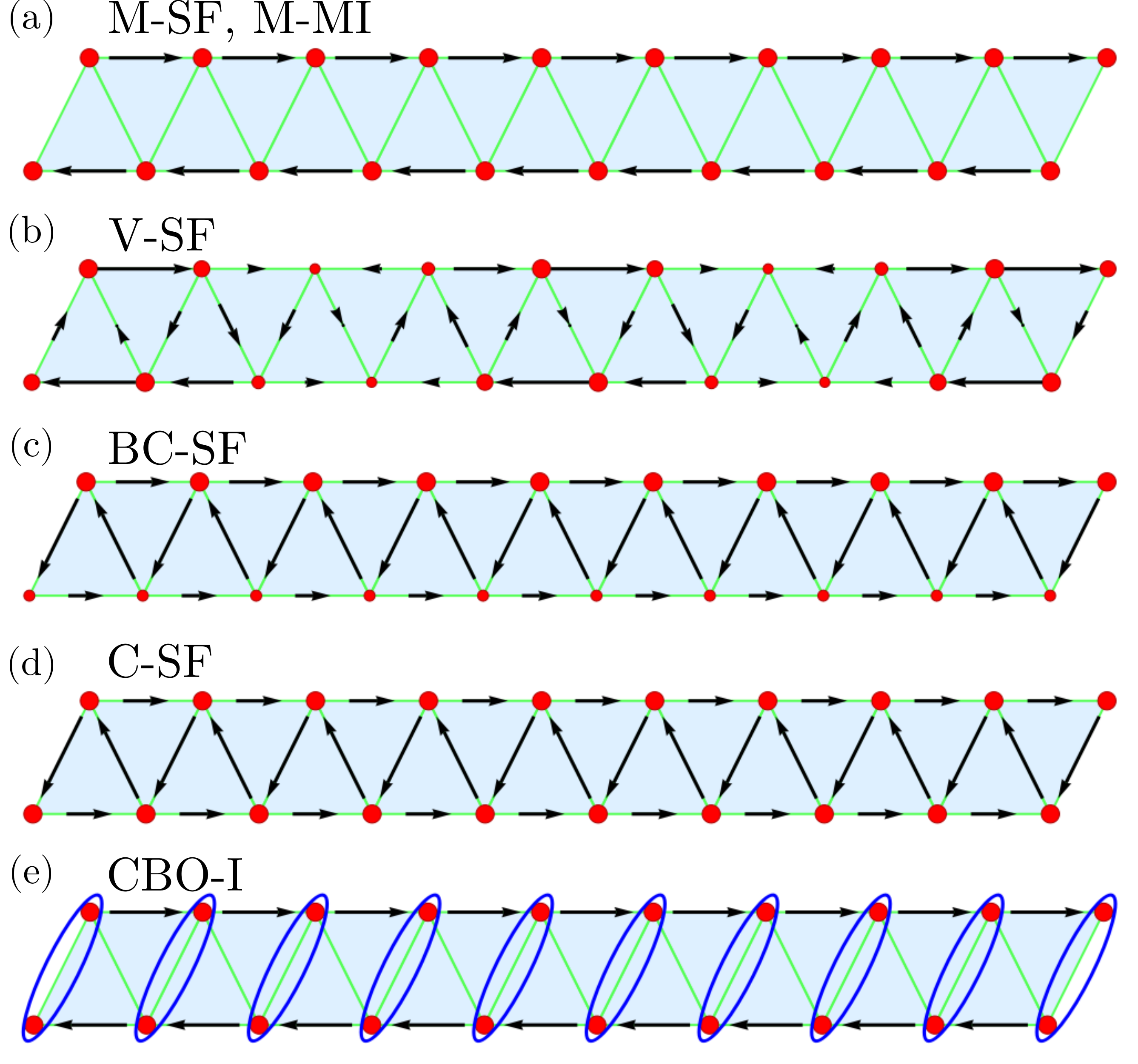}
\caption{The pattern of currents, depicted with arrows, and local densities, depicted with red disks, obtained in the numerical ground state results for the (a) Meissner states (M-SF, or M-MI) (b) vortex superfluid phases (V-SF), (c) biased chiral superfluid (BC-SF), (d) chiral superfluid (C-SF), and (e) chiral bond order insulator (CBO-I), where we also marked the bond ordering on every second rung.
We note that the local currents are not normalized to the same value for the different phases represented in the four sketches.
 }
\label{fig:currents_pattern}
\end{figure}

In this section, we focus for the case in which we have one bosonic atom every two sites, $\rho=0.5$. In Fig.~\ref{fig:phasediag_n05} we sketch the phase diagram we obtain from our numerical and analytical results which we detail in the following. 
In particular, we focus on several regions on the phase diagram. We investigate the limit of small $J_\|$ (see Sec.~\ref{sec:singlechain}), where we obtain a Meissner superfluid (M-SF). At large $J_\|$ we observe a phase transition between the Meissner superfluid (M-SF) and a vortex superfluid (V-SF) state (see Sec.~\ref{sec:twochain}). At $\chi=\pi$ a transition between a superfluid and a chiral superfluid state is present (Sec.~\ref{sec:2chain_pi}), and for $\chi\lesssim\pi$ the chiral superfluid extends to a biased chiral superfluid phase (BC-SF).
Throughout this section the value of the on-site interaction is $U/J=2.5$.

\subsection{Small $J_\|$ limit - single chain limit \label{sec:singlechain}}

\begin{figure}[!hbtp]
\centering
\includegraphics[width=.48\textwidth]{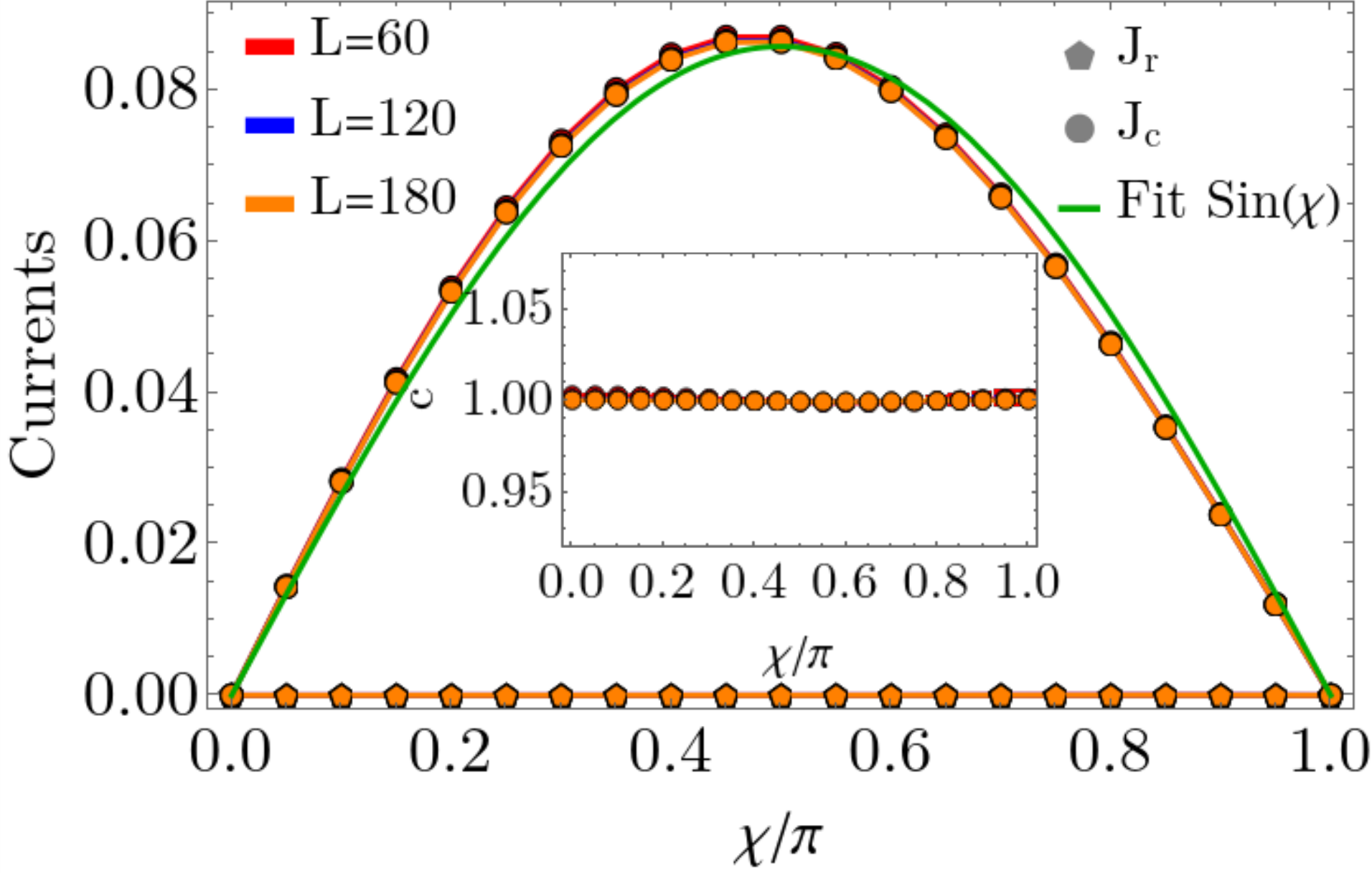}
\caption{Numerical ground state results for the average rung current, $J_r$, and the chiral current, $J_c$, as a function of the flux  $\chi$ for $J_\|/J=0.2$, $U/J=2.5$, $\rho=0.5$, $L\in\{60,120,180\}$. The inset contains the dependence of the central charge, $c$, on the flux.
We can identify the Meissner superfluid phase.
We note that the markers corresponding to the smaller system sizes are below the ones for $L=180$.
The maximal bond dimension used was $m=500$ for $L=60$, $m=900$ for $L=120$ and $m=1500$ for $L=180$.
 }
\label{fig:plots_msf}
\end{figure}

In the regime of small $J_\|/J$ it is useful to rewrite the Hamiltonian given in (\ref{eq:Hamiltonian}) as a single chain with long range complex hopping
\begin{align} 
\label{eq:Hamiltonian_chain}
 H_\text{chain}= & -J_\| \sum_{j} \left[ e^{i(-1)^j\chi}b^\dagger_{j}b_{j+2}+ \text{H.c}. \right]  \\
& -J \sum_j \left( b^\dagger_{j}b_{j+1} + \text{H.c}. \right) \nonumber \\
&+\frac{U}{2} \sum_{j} n_{j}(n_{j}-1). \nonumber
\end{align}

In this case the bosonized Hamiltonian is 
\begin{align} 
\label{eq:Hamiltonian_chain_boson}
 H_\text{chain}= & \int\frac{dx}{2\pi}\Big\{\left[uK+16\pi \rho J_\| \cos(\chi)\right]\partial_x\theta(x)^2 \\
 &\qquad+\frac{u}{K}\partial_x\phi(x)^2\Big\} \nonumber\\ 
 &+\rho^2 U\int dx \cos\left[2p\phi(x)\right], \nonumber
\end{align}
with the velocity $u$, Luttinger parameter $K$, and $p=1$ for $\rho=1$ and $p=2$ for $\rho=0.5$. For the atomic density considered in this section we expect that the interaction term does not dominate and we obtain a Luttinger liquid for which the Luttinger parameter depends on the flux $\chi$. As we see in the following this is in agreement with our numerical results. However, we note slight deviations from the analytical expectation of the dependence of effective Luttinger parameter on $\chi$ and the numerical results (see Appendix $\ref{app:K_1chain}$).

\begin{figure}[!hbtp]
\centering
\includegraphics[width=.48\textwidth]{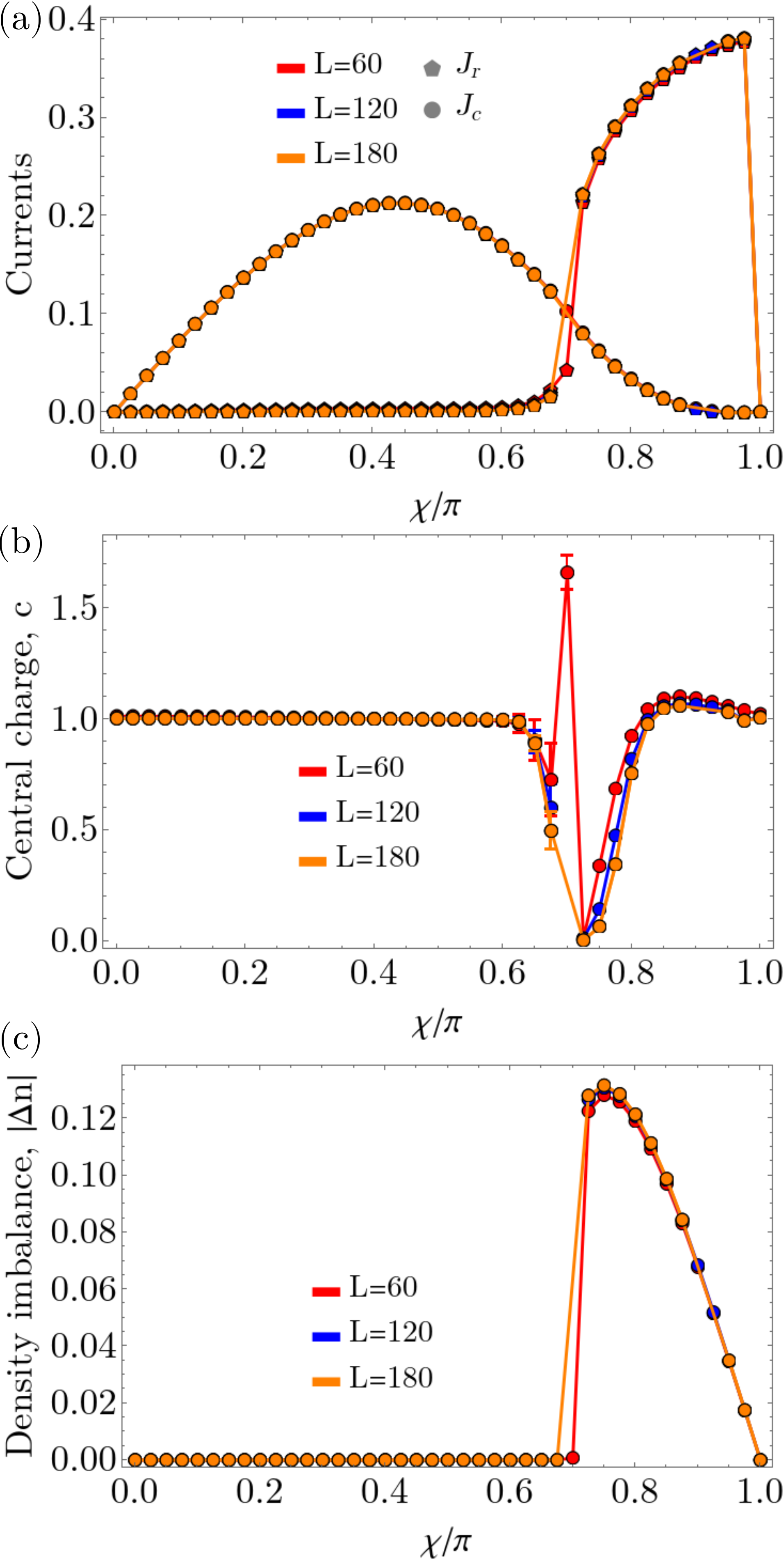}
\caption{Numerical ground state results for (a) the average rung current, $J_r$, and the chiral current, $J_c$, (b) the central charge, $c$, and (c) the absolute value of the density imbalance, $|\Delta n|$, as a function of the flux $\chi$ for $J_\|/J=0.5$, $U/J=2.5$, $\rho=0.5$, $L\in\{60,120,180\}$. We observe a transition from the Meissner superfluid to the biased chiral superfluid at $\chi\approx0.75\pi$. 
The maximal bond dimension used was $m=600$ for $L=60$, $m=1200$ for $L=120$ and $m=1800$ for $L=180$.
 }
\label{fig:plots_msf_blp}
\end{figure}

In the single chain limit the local current observables (\ref{eq:cur}) can be rewritten as
\begin{align} 
\label{eq:cur_chain}
j^\perp_j&=-iJ\left( b^\dagger_{j}b_{j+1} - \text{H.c}. \right) \\
j^\|_j&=-iJ_\|\left[e^{(-1)^j i \chi} b^\dagger_{j}b_{j+2} - \text{H.c}. \right]. \nonumber
\end{align}
In terms of the bosonic field the currents read
\begin{align} 
\label{eq:cur_chain_bos1}
j^\perp_j&=2\rho J \sin\left(\partial_x\theta\right), \\
j^\|_j&=2\rho J_\| \sin\left[2\left(\partial_x\theta+(-1)^j\frac{\chi}{2}\right)\right], \nonumber \\
j^\|_{j+1}-j^\|_j &= 2 \rho J_\| \cos\left(2\partial_x\theta\right)\sin(\chi). \nonumber
\end{align}
In the obtained gapless phase the expectation value of the rung currents will average to zero, and the chiral current has a finite value $J_c\propto\sum_j \left\langle j^\|_{j,1} - j^\|_{j,2} \right\rangle\propto \sin(\chi)$. These results are consistent with the currents expected in the Meissner superfluid phase, which are depicted in Fig.~\ref{fig:currents_pattern}(a), and their values are shown in Fig.~\ref{fig:plots_msf}.

In Fig.~\ref{fig:plots_msf} at small values of the leg tunneling amplitude, $J_\|/J=0.2$, we observe that the currents on the rungs are close to zero and the chiral current, $J_c$, has a finite value stable with increasing the system size. The central charge is $c\approx 1$ for all values of the flux, implying the existence of one gapless mode. Furthermore, in this phase, the single particle correlations decay algebraically with the distance (see Appendix $\ref{app:K_1chain}$).
Based on these considerations we can identify the Meissner superfluid state. 
However, we identify some small deviations from the expected $\sin(\chi)$ dependence of the chiral current (Fig.~\ref{fig:plots_msf}).

We can observe in Eq.~(\ref{eq:Hamiltonian_chain_boson}) for large values of $ J_\|$ and $\chi\approx\pi$ the coefficient of the first term in the Hamiltonian will vanish and eventually become negative. This instability in our bosonized model could signal a phase transition.
In the numerical results for $J_\|/J=0.5$, presented in Fig.~\ref{fig:plots_msf_blp}, we see above $\chi\gtrsim0.75$ a phase with strong currents and central charge $c\approx1$. 
Furthermore, a finite density imbalance between the two legs of the ladder is present. 
We associate this regime with the biased chiral superfluid phase. We describe in more details the nature of this phase in Sec.~\ref{sec:close_to_pi}.

We observe in the numerical results that the value of the Luttinger parameter extracted from the algebraic decay of the single particle correlations decreases and it is close to zero as we increase $\chi$ towards the phase transition between the Meissner superfluid and the biased chiral superfluid (see Appendix \ref{app:K_1chain}).

\subsection{Large $J_\|$ limit - two coupled chains limit \label{sec:twochain}}

We bosonize the Hamiltonian of the two coupled chains (\ref{eq:Hamiltonian}) in the limit where tunneling $J_\|$ along the two chains dominate. In this regime we have a pair of bosonic fields for each leg of the ladder, 
\begin{align} 
\label{eq:Hamiltonian_2chains_boson}
 H_\text{2~chains}= & \int\frac{dx}{2\pi}\left\{uK\left[\partial_x\theta_1(x)-\chi\right]^2+\frac{u}{K}\partial_x\phi_1(x)^2\right\} \nonumber\\
 +& \int\frac{dx}{2\pi}\left\{uK\left[\partial_x\theta_2(x)+\chi\right]^2+\frac{u}{K}\partial_x\phi_2(x)^2\right\} \nonumber \\
 +&\rho^2 U\sum_m\int dx \cos\left[2p\phi_m(x)\right] \\
 -& 4\rho J\int dx \cos\left[\theta_1(x)-\theta_2(x)\right]. \nonumber
\end{align}
In the following, we rewrite the Hamiltonian in terms of the symmetric and antisymmetric combinations of the bosonic fields, $\phi_{s(a)}=\frac{1}{\sqrt{2}}(\phi_1\pm\phi_2)$ and $\theta_{s(a)}=\frac{1}{\sqrt{2}}(\theta_1\pm\theta_2)$,
\begin{align} 
\label{eq:Hamiltonian_2chains_boson_2}
 H_\text{2~chains}= & \int\frac{dx}{2\pi}\left[uK\partial_x\theta_s(x)^2+\frac{u}{K}\partial_x\phi_s(x)^2\right] \\
 +& \int\frac{dx}{2\pi}\left\{uK\left[\partial_x\theta_a(x)-\sqrt{2}\chi\right]^2+\frac{u}{K}\partial_x\phi_a(x)^2\right\} \nonumber \\
 +&\rho^2 U\int dx \cos\left[\sqrt{2}p\phi_s(x)\right] \cos\left[\sqrt{2}p\phi_a(x)\right] \nonumber\\
 -& 4\rho J\int dx \cos\left[\sqrt{2}\theta_a(x)\right]. \nonumber 
\end{align}

\begin{figure}[!hbtp]
\centering
\includegraphics[width=.48\textwidth]{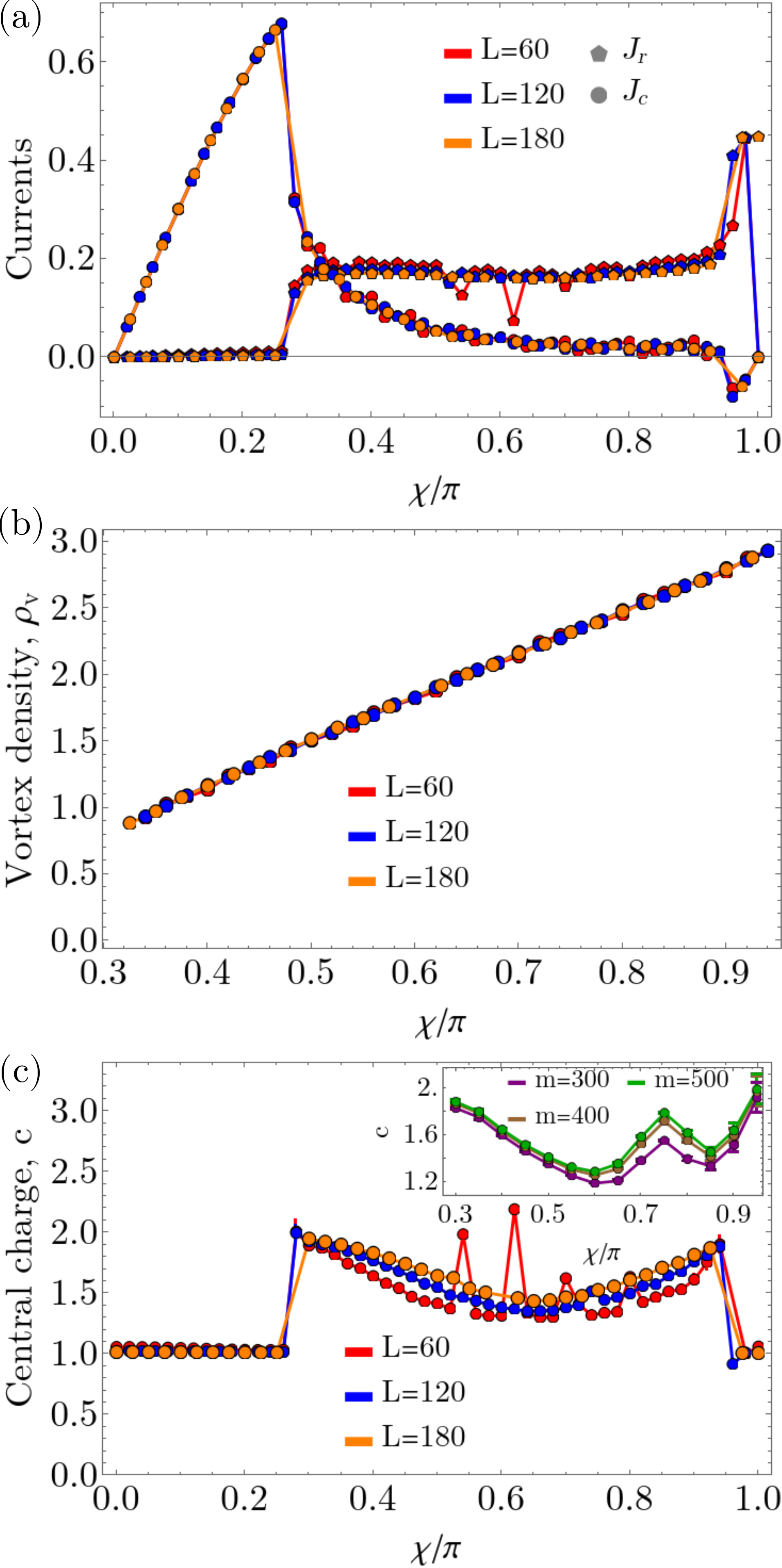}
\caption{Numerical ground state results for (a) the average rung current, $J_r$, and the chiral current, $J_c$, (b) vortex density, $\rho_v$ and (c) the central charge, $c$,  as a function of the flux $\chi$ for $J_\|/J=2$, $U/J=2.5$, $\rho=0.5$, $L\in\{60,120,\}$.
We observe a transition from the Meissner superfluid to the vortex superfluid around $\chi\approx0.27\pi$ and a transition to biased chiral superfluid at $\chi\approx0.95\pi$.
The maximal bond dimension used was $m=600$ for $L=60$, $m=1100$ for $L=120$ and $m=1800$ for $L=180$. In the inset of panel (c) we plot the central charge in the vortex superfluid regime for $L=60$ and different bond dimensions $m\in\{300,400,500\}$.
 }
\label{fig:plots_msf_vsf}
\end{figure}

Similarly the local currents (\ref{eq:localcur}) can be written as
\begin{align} 
\label{eq:cur_chain_bos2}
j^\perp_j&=2\rho J \sin\left(\sqrt{2}\theta_a\right), \\
j^\|_{j,1}-j^\|_{j,2} &= 2 \rho J_\| \cos\left(\frac{1}{\sqrt{2}}\partial_x\theta_s\right)\sin(\frac{1}{\sqrt{2}}\partial_x\theta_a-\chi). \nonumber
\end{align}

In the case of half filling, $\rho=N/(2L)=0.5$, considered in this section, we do not have commensurability effects induced by the interaction term [third line in Eq.~(\ref{eq:Hamiltonian_2chains_boson_2})]. 
We can observe in the sketch of the phase diagram, Fig.~\ref{fig:phasediag_n05}, that at $J_\|/J=2$ and $\chi\approx0.25\pi$ we have a transition between the Meissner superfluid and a vortex superfluid phase.
This phase transition can be understood by analyzing the relevance of the term $J\cos\left[\sqrt{2}\theta_a(x)\right]$ in the Hamiltonian (\ref{eq:Hamiltonian_2chains_boson_2}).
We note that we describe the nature of the biased chiral superfluid phase occurring for small $\pi-\chi$ separately in Sec.~\ref{sec:close_to_pi}. 

When the term  $J\cos\left[\sqrt{2}\theta_a(x)\right]$ dominates, we have a gapless symmetric mode and a gapped antisymmetric mode for which the field is pinned to the minima of the potential, $\theta_a=0$.
In this case the rung currents vanish and the chiral current $J_c\propto \sin(\chi)$, thus, this corresponds to the Meissner superfluid. This is in agreement with the numerical results at presented in Fig.~\ref{fig:plots_msf_vsf}(a) for the low values of $\chi$.
The central charge in this regime is $c=1$, as seen in Fig.~\ref{fig:plots_msf_vsf}(c).

If both the symmetric and antisymmetric sectors are gapless we have the vortex superfluid. The vortex phase exhibits incommensurate currents patterns, as, for example, depicted in Fig.~\ref{fig:currents_pattern}(b). 
One can identify this phase for $0.25\lesssim\chi\lesssim 0.95$, as seen from the values of the average of the rung and chiral currents in Fig.~\ref{fig:plots_msf_vsf}(a).
We extract numerically the vortex density by performing the Fourier transform of the space dependence of the rung currents and obtaining the periodicity of the vortices. The vortex density is plotted in Fig.~\ref{fig:plots_msf_vsf}(b) as a function of the flux, which shows a linear behavior.
This dependence can be understood in bosonization by computing the rung current-current correlations
\begin{align} 
\label{eq:cur_cur_corr}
\langle j^\perp(x) j^\perp(0) \rangle&=4 \rho^2 J^2 
\left\langle \sin\left[\sqrt{2}\theta_a(x)\right]\sin\left[\sqrt{2}\theta_a(0)\right] \right\rangle\nonumber \\
&=2 \rho^2  J^2 \left\langle\cos\left[\sqrt{2}(\theta_a(x)-\theta_a(0))\right]\right\rangle \\
&\propto x^{-1/(4K)} \cos(\chi x), \nonumber
\end{align}
where $x$ is the lattice position in the continuum limit. The frequency of the oscillations gives the expected vortex density, in agreement with the values obtained numerically in Fig.~\ref{fig:plots_msf_vsf}(b).
We note that we do not observe any commensurate vortex densities $\rho_v$, as obtained for the square ladder \cite{OrignacGiamarchi2001}.

We can observe that the finite size effects are more prominent in the vortex phase, both in the currents [Fig.~\ref{fig:plots_msf_vsf}(a)] and the central charge [Fig.~\ref{fig:plots_msf_vsf}(c)]. However, as we increase the system size the value of the central charge becomes closer to the expected value of $c=2$, similarly in the inset of Fig.~\ref{fig:plots_msf_vsf}(c) we analyze the impact of the numerical bond dimension used in the MPS representation.
The behavior of the system for $\chi$ close to and equal to $\pi$ is discussed in the following sections.

\subsection{Fully frustrated ladder at $\chi=\pi$ \label{sec:2chain_pi}}

In the following, we analyze the behavior in the case of $\chi=\pi$. In the phase diagram this corresponds to a particular line, marked in Fig.~\ref{fig:phasediag_n05}. In this case the Hamiltonian (\ref{eq:Hamiltonian}) becomes
\begin{align} 
\label{eq:Hamiltonian_pi}
 H=&  J_\| \sum_j \left( b^\dagger_{j,1}b_{j+1,1}+ b^\dagger_{j,2}b_{j+1,2} + \text{H.c}. \right) \\
 -&J \sum_j \left( b^\dagger_{j,1}b_{j,2}+b^\dagger_{j+1,1}b_{j,2} + \text{H.c}. \right) \nonumber \\
+&\frac{U}{2} \sum_{j} n_{j,m}(n_{j,m}-1). \nonumber
\end{align}
We employ the transformation $b_{j,m}\to (-1)^j b_{j,m}$ to change the sign of the kinetic energy along the legs and obtain
\begin{align} 
\label{eq:Hamiltonian_pi_transformed}
 H=-& J_\| \sum_j \left( b^\dagger_{j,1}b_{j+1,1}+ b^\dagger_{j,2}b_{j+1,2} + \text{H.c}. \right) \\
 -&J \sum_j \left( b^\dagger_{j,1}b_{j,2}-b^\dagger_{j+1,1}b_{j,2} + \text{H.c}. \right) \nonumber \\
+&\frac{U}{2} \sum_{j} n_{j,m}(n_{j,m}-1). \nonumber
\end{align}
In this representation, the hopping on the rungs has an alternating sign. Similar models have been investigated in Refs.~\cite{MishraParamekanti2013, Greschner_2013_zigzagbosons}.
The alternating rung hopping terms cancels the lowest order term which we considered in the previous sections, such that we have to take into account the next order terms of the expansion in the bosonic fields
\begin{align} 
\label{eq:hopping_gradients}
b^\dagger_{j,1}b_{j,2}&=\rho e^{-i \left[ \theta_1 (aj)-\theta_2 \left(aj\right)-\frac{a}{2}\partial_x\theta_2(aj) \right]},  \\
b^\dagger_{j+1,1}b_{j,2}&=\rho e^{-i \left[\theta_1 (aj)-\theta_2 \left(aj\right)+\frac{a}{2}\partial_x\theta_1(aj)\right]},&\nonumber
\end{align}
where the gradients stem from the triangular geometry of the ladder.
We obtain for the tunneling along the rungs
\begin{align} 
\label{eq:Hamiltonian_perp_pi}
 H_\perp&=-2\rho J\int dx \Big\{\cos\left[\theta_1 (x)-\theta_2 \left(x\right)-\frac{1}{2}\partial_x\theta_2(x)\right]\nonumber\\
 &\quad-\cos\left[\theta_1 (x)-\theta_2 \left(x\right)+\frac{1}{2}\partial_x\theta_1(x)\right]\Big\} \\
 &=-2\rho J\int dx \sin\left(\sqrt{2}\theta_a+ \frac{1}{\sqrt{2}}\partial_x\theta_a\right)\sin\left(\frac{1}{\sqrt{2}}\partial_x\theta_s\right) \nonumber\\
 &\approx -\sqrt{2}\rho J\int dx \sin\left(\sqrt{2}\theta_a\right) \partial_x \theta_s,\nonumber
\end{align}
where in the last line we expanded the sine and neglected the contributions stemming from $\partial_x\theta_a$. In the regime $H_\perp$ dominates, one obtains ground states which break the $\mathbb{Z}_2$ symmetry of the model, for example obtaining a spin nematic phase in a triangular spin ladder \cite{Nersesyan_1998_zigzagspinladder}, or a chiral superfluid in a bosonic zig-zag ladder \cite{Greschner_2013_zigzagbosons}.

The coupling between $\theta_s$ and $\theta_a$ present in $H_\perp$ can be analyzed with the help of a self-consistent mean-field approach \cite{Nersesyan_1998_zigzagspinladder}. We outline this approach in Appendix~\ref{app:mf_cblp}. We obtain two solutions for the ground state, which break the $\mathbb{Z}_2$ symmetry, for which the field $\theta_a$ is fixed either to $\frac{\pi}{2\sqrt{2}}$, or to $\frac{3\pi}{2\sqrt{2}}$, and for both $\left\langle \partial_x \theta_s(x)\right\rangle$ has a finite value.

\begin{figure}[!hbtp]
\centering
\includegraphics[width=.48\textwidth]{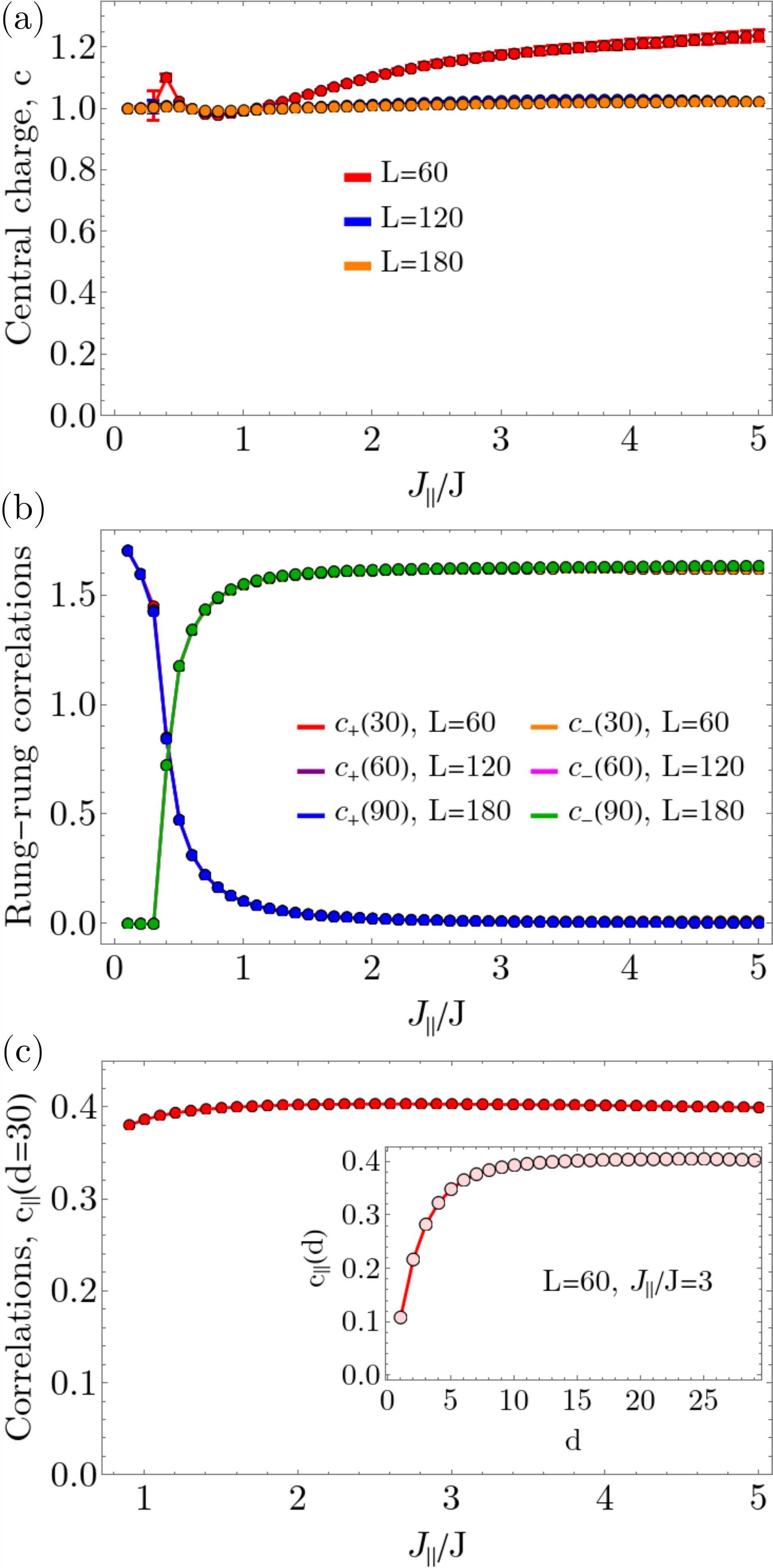}
\caption{Numerical ground state results for (a) the central charge, (b) the saturation value of the rung-rung correlations $c_\pm$ (\ref{eq:corr_pi}), (c) the saturation value of the current-current correlations on the legs $c_\|$ (\ref{eq:corr_pi_2}), as a function of $J_\|$, for $\chi=\pi$, $U/J=2.5$, $\rho=0.5$, $L\in\{60,120,180\}$.
In the inset of panel (c) we show the distance dependence of $c_\|(d)$ at $J_\|/J=3$.
We observe a transition from the superfluid to the chiral superfluid at $J_\|/J\approx 0.35$.
The maximal bond dimension used was $m=500$ for $L=60$, $m=1000$ for $L=120$ and  $m=1500$ for $L=180$.
 }
\label{fig:line_pi}
\end{figure}
 
Using the results of the mean-field theory, we can compute the currents present in the triangular ladder
\begin{align}
    \label{eq:currents_pi}
    j^\perp_j&=-i J\left(b^\dagger_{j,1}b_{j,2}-\text{H.c.}\right) \\
    &\propto \sin\left(\sqrt{2}\theta_a\right)\cos\left(\frac{1}{4}\partial_x\theta_s\right),\nonumber \\
    j^\perp_j+j^\perp_{j+1}&=-i J\left(b^\dagger_{j,1}b_{j,2}-b^\dagger_{j+1,1}b_{j,2}-\text{H.c.}\right) \nonumber\\
    &\propto \cos\left(\sqrt{2}\theta_a\right)\sin\left(\frac{1}{\sqrt{2}}\partial_x\theta_s\right)=0.\nonumber
\end{align}
\begin{align}
    \label{eq:currents_pi_2}
    j^\|_{j,1}&=-i J_\|\left(b^\dagger_{j,1}b_{j+1,1}-\text{H.c.}\right) \\
    &\propto \sin\left(\frac{1}{\sqrt{2}}\partial_x\theta_s\right)\cos\left(\frac{1}{\sqrt{2}}\partial_x\theta_a\right)\nonumber \\
    &~~~+\cos\left(\frac{1}{\sqrt{2}}\partial_x\theta_s\right)\sin\left(\frac{1}{\sqrt{2}}\partial_x\theta_a\right)\propto \partial_x\theta_s, \nonumber \\
    j^\|_{j,2}&=-i J_\|\left(b^\dagger_{j,2}b_{j+1,2}-\text{H.c.}\right) \nonumber\\
    &\propto \partial_x\theta_s, \nonumber \\
    j^\|_{j,1}-j^\|_{j,2} &\propto \cos\left(\frac{1}{\sqrt{2}}\partial_x\theta_s\right)\sin\left(\frac{1}{\sqrt{2}}\partial_x\theta_a\right)=0.\nonumber
\end{align}
We have obtained the chiral superfluid (C-SF) phase \cite{Greschner_2013_zigzagbosons}, in which we have finite currents on the rungs, equal but with opposite directions on consecutive rungs, and finite currents on the legs flowing in the same direction on the two legs. The pattern is depicted in Fig.~\ref{fig:currents_pattern}(d).
The direction of the currents is determined by which of the two $\mathbb{Z}_2$ symmetry broken states one considers.

Thus, for $\chi=\pi$ at half-filling we expect to observe a superfluid phase at small values of $J_\|/J$ (depicted with the thick green line for $\chi=\pi$ in Fig.~\ref{fig:phasediag_n05}), and  the chiral superfluid phase (depicted with the thick dark blue line for $\chi=\pi$ in Fig.~\ref{fig:phasediag_n05}. This is in agreement with the numerical ground state results, shown in Fig.~\ref{fig:line_pi}.

\begin{figure}[!hbtp]
\centering
\includegraphics[width=.48\textwidth]{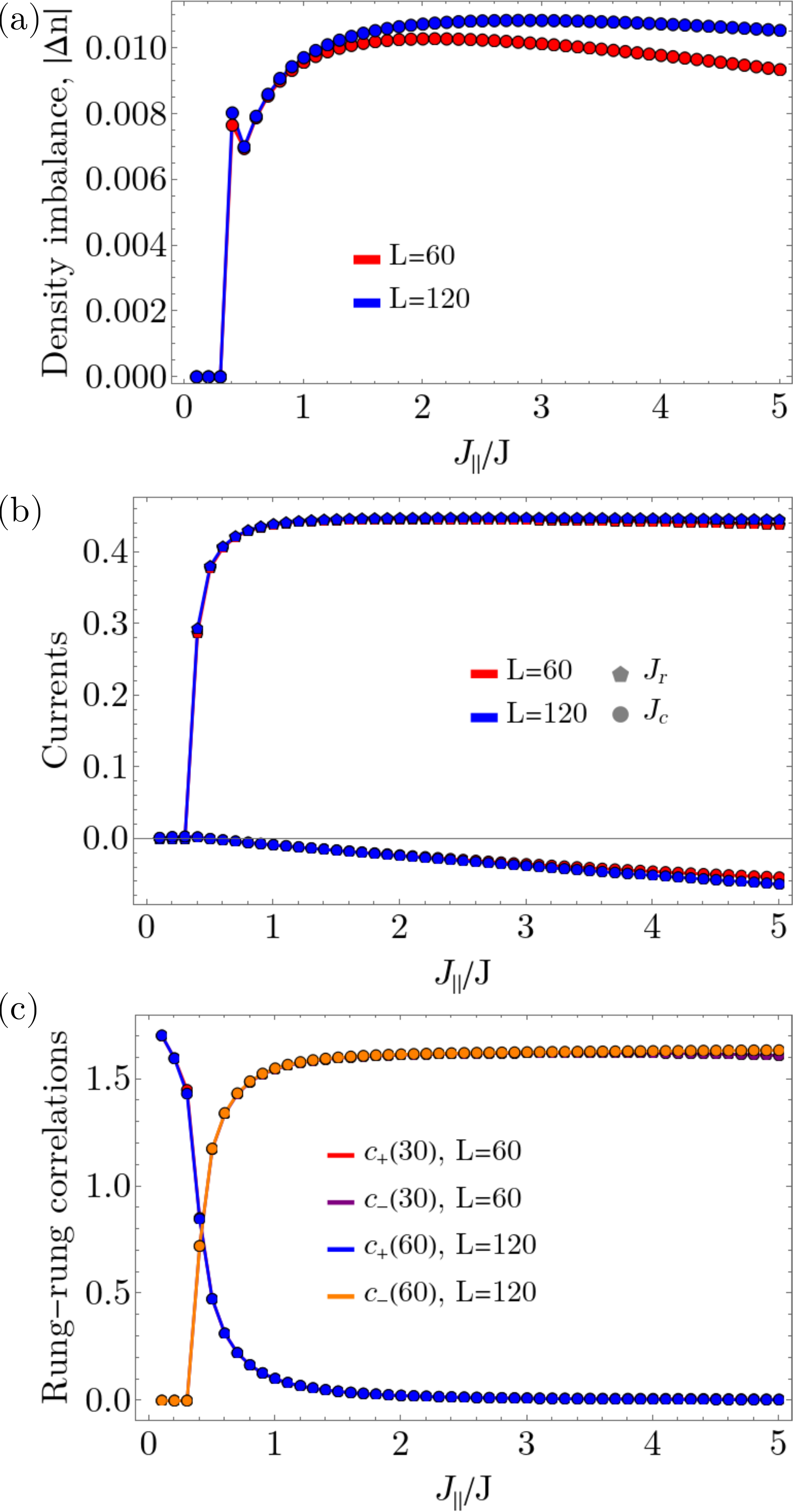}
\caption{Numerical ground state results for (a) the absolute value of the density imbalance, $|\Delta n|$, (b) the average rung current, $J_r$, and the chiral current, $J_c$, and 
(c) the saturation value of the rung-rung correlations $c_\pm$ (\ref{eq:corr_pi}), as a function of $J_\|$, for $\chi=0.99\pi$, $U/J=2.5$, $\rho=0.5$, $L\in\{60,120\}$.
We observe a transition from the Meissner superfluid to the biased chiral superfluid at $J_\|/J\approx 0.35$.
The maximal bond dimension used was $m=600$ for $L=60$ and $m=1200$ for $L=120$.
We note that the apparent non-monotonic behavior seen in $|\Delta n|$ in (a) for $J_\|/J\approx0.4$ is within the numerical errors in the regime close to the phase transition and it is not observed in the observables depicted in (b) and (c).
 }
\label{fig:line_close_to_pi}
\end{figure}

The central charge varies only slightly as a function of $J_\|$ around the expected value for the two phases, $c=1$, see Fig.~\ref{fig:line_pi}(a).
If we look at the average rung current we find that they are vanishing both in Fig.~\ref{fig:plots_msf_blp}(a) for $J_\|/J=0.5$ and in Fig.~\ref{fig:plots_msf_vsf}(a) for $J_\|/J=2$.
However, this does not imply that we cannot be in the chiral superfluid phase for these parameters. One explanation can be that in our numerical ground-state calculations we converge to an equal superposition of the two possible $\mathbb{Z}_2$ symmetry broken states (Appendix~\ref{app:mf_cblp}). This would imply that the measured expectation values of the measured local currents are zero, as the currents in the two states have the same magnitude, but a different sign.
In order to confirm this behavior we compute the following rung-rung correlations
\begin{align}
    \label{eq:corr_pi}
    c_+(d)=&\left(b^\dagger_{j,1} b_{j,2}+ b^\dagger_{j,2} b_{j,1}\right)\left(b^\dagger_{j+d,2} b_{j+d,1}+b^\dagger_{j+d,1}b_{j+d,2}\right) \nonumber\\
    &\propto \cos\left[\sqrt{2}\theta_a(x=ad)\right]\cos\left[\sqrt{2}\theta_a(0)\right],  \\
   c_-(d)=& \left(b^\dagger_{j,1} b_{j,2}- b^\dagger_{j,2} b_{j,1}\right)\left(b^\dagger_{j+d,2} b_{j+d,1}-b^\dagger_{j+d,1}b_{j+d,2}\right) \nonumber\\
   & \propto \sin\left[\sqrt{2}\theta_a(x=ad)\right]\sin\left[\sqrt{2}\theta_a(0)\right]. \nonumber
\end{align}
In Fig.~\ref{fig:line_pi}(b) one can observe that for $J_\|/J\lesssim 0.3$ we find that $c_+$ saturates at long distances ($d=30$) signaling that $\theta_a$ orders in the potential of the $\cos\left(\sqrt{2}\theta_a\right)$ term. This phase corresponds to a superfluid state as the $\chi\to\pi$ limit of the Meissner superfluid. 
When we increase $J_\|/J\gtrsim 0.3$ we see that $c_+$ decreases to a small value at the longest distance we consider and $c_-$ is the one that saturates at a large distances, implying that $\sin\left(\sqrt{2}\theta_a\right)$ gaps the antisymmetric sector, as expected for the chiral superfluid phase. 
This is further supported by the current-current correlations along the legs
\begin{align}
    \label{eq:corr_pi_2}
   c_\|(d)=&J_\|^2\left(b^\dagger_{j,m} b_{j+1,m}-b^\dagger_{j+1,m} b_{j,m}\right)\\
   &\quad\times\left(b^\dagger_{j+d,m} b_{j+d+1,m}-b^\dagger_{j+d+1,m}b_{j+d,m}\right).\nonumber
\end{align}
In order to reduce the finite size effects in the numerical calculation of $c_+(d)$, $c_-(d)$ and $c_\|(d)$ (\ref{eq:corr_pi})-(\ref{eq:corr_pi_2}) [presented in Figs.~\ref{fig:line_pi}(b)-(c), Figs.~\ref{fig:pi_to_square}(b)-(c) and Fig.~\ref{fig:line_close_to_pi}] we normalized $b_{j,m}\to b_{j,m}/\sqrt{\langle n_{j,m} \rangle}$.
The leg current-current correlations, Fig.~\ref{fig:line_pi}(c), are finite in the regime corresponding to the chiral superfluid and almost constant as a function of $J_\|$. In the inset of  Fig.~\ref{fig:line_pi}(c) we show the saturation behavior of the current-current correlations at large distances.

We note that depending on the initial states and gauge chosen in the numerical calculations, one can converge to just one of the  $\mathbb{Z}_2$ symmetry broken states and see finite values of the currents as presented in Fig.~\ref{fig:currents_pattern}(d).
We give more details in this regard in Appendix~\ref{app:dmrg_cblp}.

\subsection{Large $J_\|$ limit and small $\chi_0=\pi-\chi$. \label{sec:close_to_pi}}

In this section, we address the question if the chiral superfluid phase extends also at finite values of $\chi_0=\pi-\chi$, as we can observe for $J_\|/J=2$ in Fig.~\ref{fig:plots_msf_vsf}(c) a transition to a $c=1$ phase above $\chi\approx 0.95\pi$.
We approach this by rewriting the Hamiltonian (\ref{eq:Hamiltonian}) in terms of $\chi_0$
\begin{align} 
\label{eq:Hamiltonian_close_to_pi}
 H=&  J_\| \sum_j \left(e^{i \chi_0} b^\dagger_{j,1}b_{j+1,1}+ e^{-i \chi_0} b^\dagger_{j,2}b_{j+1,2} + \text{H.c}. \right) \nonumber\\
 -&J \sum_j \left( b^\dagger_{j,1}b_{j,2}+b^\dagger_{j+1,1}b_{j,2} + \text{H.c}. \right)  \\
+&\frac{U}{2} \sum_{j} n_{j,m}(n_{j,m}-1). \nonumber
\end{align}
Employing the transformation $b_{j,m}\to (-1)^j b_{j,m}$, as in Sec.~\ref{sec:2chain_pi} we obtain
\begin{align} 
\label{eq:Hamiltonian_close_to_pi_1}
 H=&  -J_\| \sum_j \left(e^{i \chi_0} b^\dagger_{j,1}b_{j+1,1}+ e^{-i \chi_0} b^\dagger_{j,2}b_{j+1,2} + \text{H.c}. \right) \nonumber\\
 -&J \sum_j \left( b^\dagger_{j,1}b_{j,2}-b^\dagger_{j+1,1}b_{j,2} + \text{H.c}. \right)  \\
+&\frac{U}{2} \sum_{j} n_{j,m}(n_{j,m}-1). \nonumber
\end{align}
The bosonized Hamiltonian, in terms of the symmetric and antisymmetric fields is given by
\begin{align} 
\label{eq:Hamiltonian_close_to_pi_boson}
 H= & \int\frac{dx}{2\pi}\left[uK\partial_x\theta_s(x)^2+\frac{u}{K}\partial_x\phi_s(x)^2\right] \\
 +& \int\frac{dx}{2\pi}\left\{uK\left[\partial_x\theta_a(x)+\sqrt{2}\chi_0/a\right]^2+\frac{u}{K}\partial_x\phi_a(x)^2\right\} \nonumber \\
 +&\rho^2 U\int dx \cos\left[\sqrt{2}p\phi_s(x)\right] \cos\left[\sqrt{2}p\phi_a(x)\right] \nonumber\\
 -& \sqrt{2}\rho J\int dx \sin\left[\sqrt{2}\theta_a(x)\right]\partial_x\theta_s(x). \nonumber 
\end{align}
We observe that the Hamiltonian is similar to the one obtained in Eq.~(\ref{eq:Hamiltonian_2chains_boson_2}) for the two coupled chains. However it exhibits the same coupling between the symmetric and antisymmetric sectors as in the case of $\chi=\pi$ (\ref{eq:Hamiltonian_perp_pi}). 
Thus, when both sectors are gapless one would obtain the vortex superfluid phase. In the following, we analyze the case in which the coupling $\sin\left[\sqrt{2}\theta_a(x)\right]\partial_x\theta_s(x)$ gaps the antisymmetric sector, by looking at behavior of the currents 
\begin{align}
    \label{eq:currents_pi_1}
    j^\perp_j&=-i J\left(b^\dagger_{j,1}b_{j,2}-\text{H.c.}\right)\propto\sin\left(\sqrt{2}\theta_a\right), \\
    j^\perp_j+j^\perp_{j+1}&=-iJ\left(b^\dagger_{j,1}b_{j,2}-b^\dagger_{j+1,1}b_{j,2}-\text{H.c.}\right) \nonumber \\
    &\propto \cos\left(\sqrt{2}\theta_a\right)\sin\left(\frac{1}{\sqrt{2}}\partial_x\theta_s\right)=0.\nonumber \\
     j^\|_{j,1}-j^\|_{j,2}&=-2\rho J_\| 
     \cos\left(\frac{1}{\sqrt{2}}\partial_x\theta_s\right)\sin\left(\frac{1}{\sqrt{2}}\partial_x\theta_a+\chi_0\right) \nonumber \\
    &\propto\sin(\chi_0)=\sin(\chi).  \nonumber
\end{align}

In this situation, we obtain two degenerate ground states breaking the $\mathbb{Z}_2$ symmetry of the ladder, with a similar pattern of currents as in the case of $\chi=\pi$ for the chiral superfluid (\ref{eq:currents_pi})-(\ref{eq:currents_pi_2}). 
However, at finite $\chi_0$ the currents on the legs have different values, even though they flow in the same direction. The pattern of the currents is depicted in Fig.~\ref{fig:currents_pattern}(c).
As the currents on the two consecutive rungs are equal, this implies that the densities on the two legs of the ladder have to be different, due to the continuity relation between densities and currents.
Thus, there exists a density imbalance between the two legs of the ladder. We label this phase as the biased chiral superfluid. 
We note that the biased chiral superfluid phase is not the direct equivalent of the biased ladder phase observed in square ladders \cite{WeiMueller2014, GreschnerVekua2016}, as in the biased phase of the square ladder the currents resemble the ones in the Meissner phase. In the considered model, due to the geometry of the triangular ladder, we obtain a biased phase with similar currents as in the chiral superfluid.

Numerically, we analyze the case of $\chi=0.99\pi$ as a function of the value of $J_\|/J$, Fig.~\ref{fig:line_close_to_pi}. Above $J_\|/J\approx0.35$ we observe that the density imbalance becomes finite [see Fig.~\ref{fig:line_close_to_pi}(a)]. This fact together with the finite currents, Fig.~\ref{fig:line_close_to_pi}(b), point towards the biased chiral superfluid. 
The behavior of the rung-rung correlations shown in Fig.~\ref{fig:line_close_to_pi}(c) is very similar to the one observed for $\chi=\pi$ in Fig.~\ref{fig:line_pi}.
The computed observables have a weak dependence as we increase $J_\|/J$ above the transition threshold, thus, we can be confident that the imbalanced phase observed at small values of $J_\|$ in Sec.~\ref{sec:singlechain} has the same nature as at large $J_\|$.
We discuss the numerical challenges in converging in the degenerate ground state manifold in Appendix~\ref{app:dmrg_cblp}.

\subsection{Transition to the square ladder\label{sec:2chain_square}}

\begin{figure}[!hbtp]
\centering
\includegraphics[width=.48\textwidth]{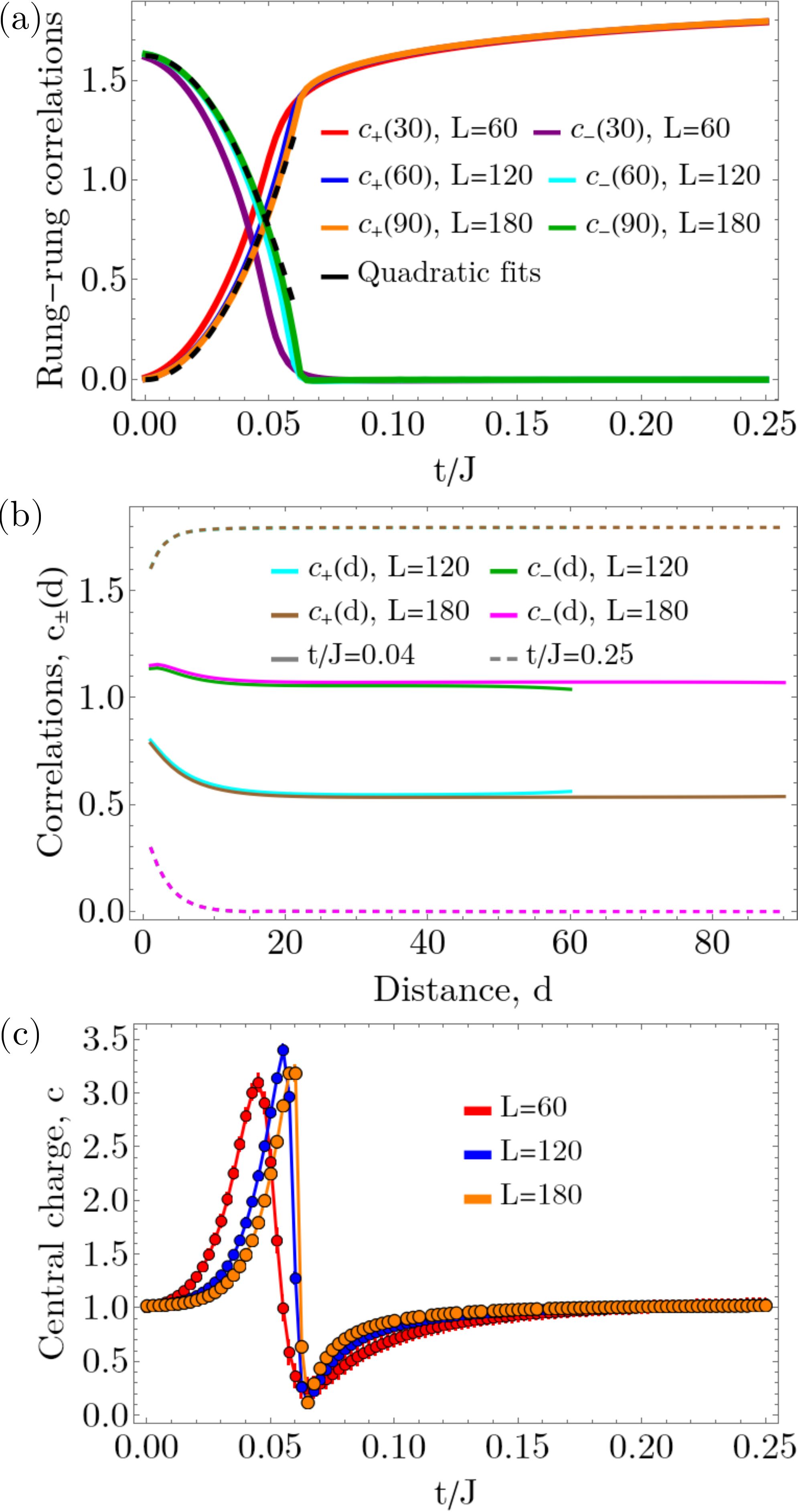}
\caption{Numerical ground state results for (a) the saturation value of the rung-rung correlations $c_\pm$, (\ref{eq:corr_pi}) as a function of the parameter $t$ interpolating between the triangular ladder at $\chi=\pi$ and the square ladder without a flux. The black lines depict the fit of the dependence on $t$ as given in Eq.~(\ref{eq:sq_corr}).
(b) The decay of the rung-rung correlations $c_\pm$ as a function of distance for $t/J\in\{0.04, 0.25\}$ and $L\in\{120,180\}$.
Note that the continuous lines correspond to  $t/J=0.04$ and the dashed lines to $t/J=0.25$.
(c) The central charge, $c$, as a function of $t/J$.
The parameter used are $J_\|/J=5$, $U/J=2.5$, $\rho=0.5$, $L\in\{60,120,180\}$. The maximal bond dimension used was $m=500$ for $L=60$, $m=1000$ for $L=120$ and $m=1500$ for $L=180$.
 }
\label{fig:pi_to_square}
\end{figure}

An interesting extension of model for the triangular ladder at $\chi=\pi$ (\ref{eq:Hamiltonian_pi_transformed}) is the interpolation towards the  square ladder without a flux, as in the following
\begin{align} 
\label{eq:Hamiltonian_pi_square}
 H=-& J_\| \sum_j \left( b^\dagger_{j,1}b_{j+1,1}+ b^\dagger_{j,2}b_{j+1,2} + \text{H.c}. \right)  \\
 -&J \sum_j \left( b^\dagger_{j,1}b_{j,2} + \text{H.c}. \right) \nonumber \\
 +&(J-t) \sum_j \left(b^\dagger_{j+1,1}b_{j,2} + \text{H.c}. \right) \nonumber \\
+&\frac{U}{2} \sum_{j} n_{j,m}(n_{j,m}-1), \nonumber
\end{align}
for $t=0$ we recover Eq.~(\ref{eq:Hamiltonian_pi_transformed}) and for $t=J$ we obtain a square ladder.

In the bosonized language we obtain for the tunneling along the rungs
\begin{align} 
\label{eq:Hamiltonian_perp_square}
H_\perp=& -\sqrt{2}\rho J \int dx \sin\left(\sqrt{2}\theta_a\right) \partial_x \theta_s\\
&-2\rho t \int dx \cos\left(\sqrt{2}\theta_a\right). \nonumber
\end{align}
This model allows us to investigate the existence of the mean-field solutions in the case in which the field $\theta_a$ is not pinned in the minima of the potential $\sin\left(\sqrt{2}\theta_a\right)$, but at a different value due to the presence of the additional term $\cos\left(\sqrt{2}\theta_a\right)$. We show how one can perform the mean-field approach for this situation in Appendix~\ref{app:mf_cblp}.

In the case of $t=0$ we saw in the previous section that at the large value of $J_\|/J$ the field $\theta_a$ is pinned to the minima of the potential $\pm\sin(\sqrt{2}\theta_a)$. In the limit of $t/J\to 1$, as we get closer to the square ladder, the field $\theta_a$ will be pinned to the minima of the potential $\cos(\sqrt{2}\theta_a)$.
However, in between these two values we can expect a chiral superfluid regime in which $\langle\theta_a\rangle$ corresponds to the minima of the sum of the two potentials, as described in Appendix~\ref{app:mf_cblp}, and a transition to a superfluid state for which the field is pinned to $\theta_a=0$.
Based on the mean-field approach we can analyze the behavior of the saturation value of the correlations $c_-$ and $c_+$ as a function of $t$ (see Appendix~\ref{app:mf_cblp})
\begin{align} 
\label{eq:sq_corr}
&c_-(d\to\infty)\propto a_-- b_- t^2,\\
&c_+(d\to\infty)\propto b_+ t^2,\nonumber
\end{align}
where $a_-$, $b_-$ and $b_+$ are constant which can depend on the other parameters of the model.
In Fig.~\ref{fig:pi_to_square}(a) we observe that for $t\lesssim 0.07$ the scaling of the saturation value of the correlations $c_-$ and $c_+$ as a function of $t$ is in agreement with the results of the mean-field theory (\ref{eq:sq_corr}). This shows that the value of $\langle\theta_a\rangle$ depends on the value of $t$.
Above $t\gtrsim 0.07$, only $c_+$ saturates to a finite value at large distances which means that the term $\cos(\sqrt{2}\theta_a)$ dominates and $\langle\theta_a\rangle=0$.
The saturation behavior of the correlations $c_\pm$ as a function of  distance in the two regimes is shown in Fig.~\ref{fig:pi_to_square}(b).
The central charge, Fig.~\ref{fig:pi_to_square}(c), has values $c\approx 1$ in both phase and strong variation close to the phase transition. However, the region in which it deviates from the expected $c=1$ value becomes smaller for the larger system sizes.

\section{Phase diagram of hardcore bosons, $\rho=0.5$ \label{sec:hardcore}}

\begin{figure}[!hbtp]
\centering
\includegraphics[width=.48\textwidth]{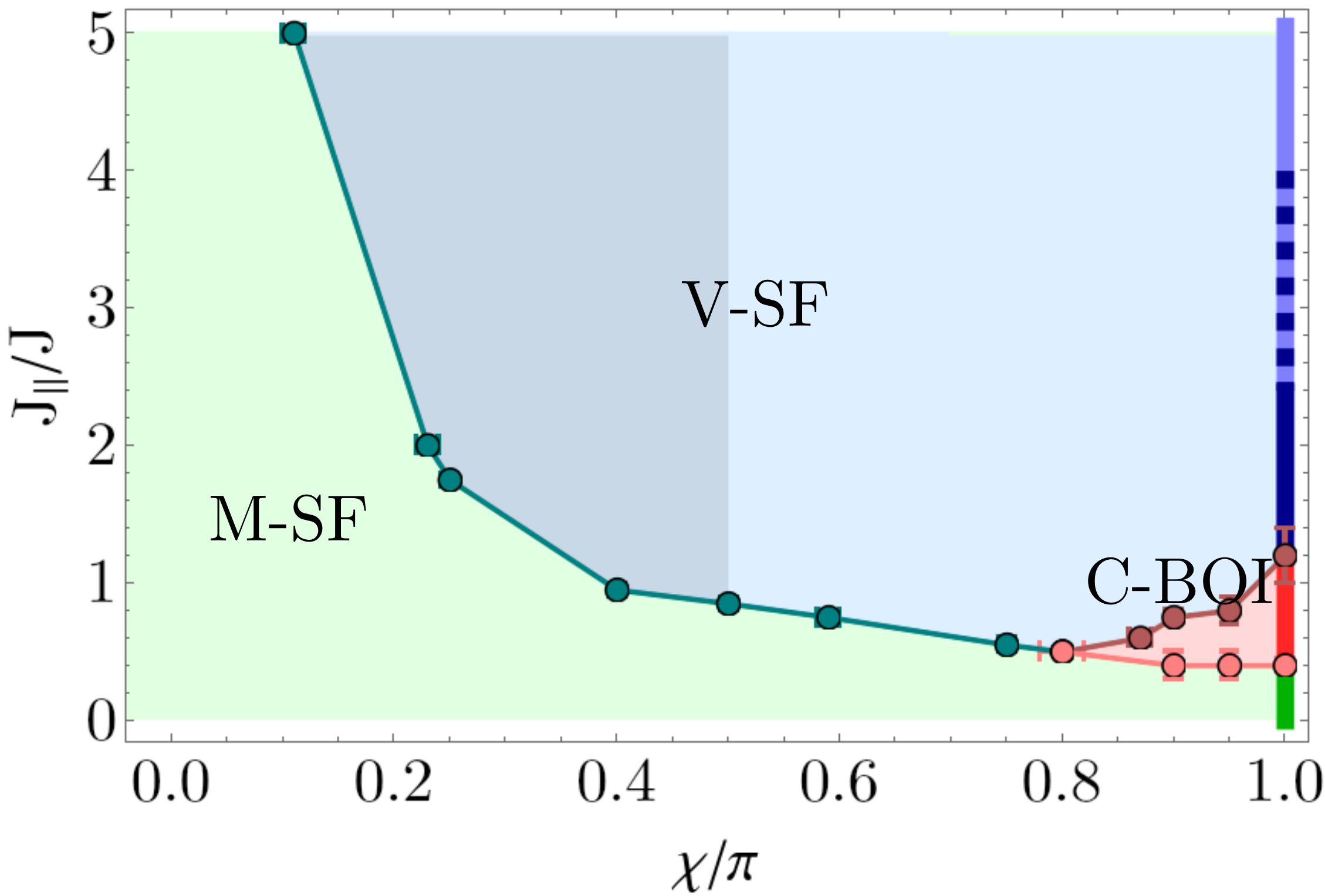}
\caption{Sketch of the phase diagram for hardcore bosons at half filling, $\rho=0.5$, $U/J=\infty$.
The identified phases (see text) are the the Meissner superfluid (M-SF), the vortex superfluid (V-SF) and chiral bond order insulator (C-BOI).
In the shaded area in the vortex phase, which extends up to $
\chi=0.5\pi$, we identified also a commensurate frequency in the behavior of the currents.
At $\chi=\pi$ (marked by a thick vertical line) we observe the following phases: a superfluid phase (green), a bond order insulator (red), a chiral superfluid phase (dark blue), a two mode superfluid (blue). The phase transition between the chiral superfluid and the two mode superfluid shows large finite size effects in the dashed region [see the explanation in the main text, Fig.~\ref{fig:hardcore_pi}(b) and Appendix~\ref{app:dmrg_cblp}].
 }
\label{fig:phasediag_hardcore}
\end{figure}

\begin{figure}[!hbtp]
\centering
\includegraphics[width=.48\textwidth]{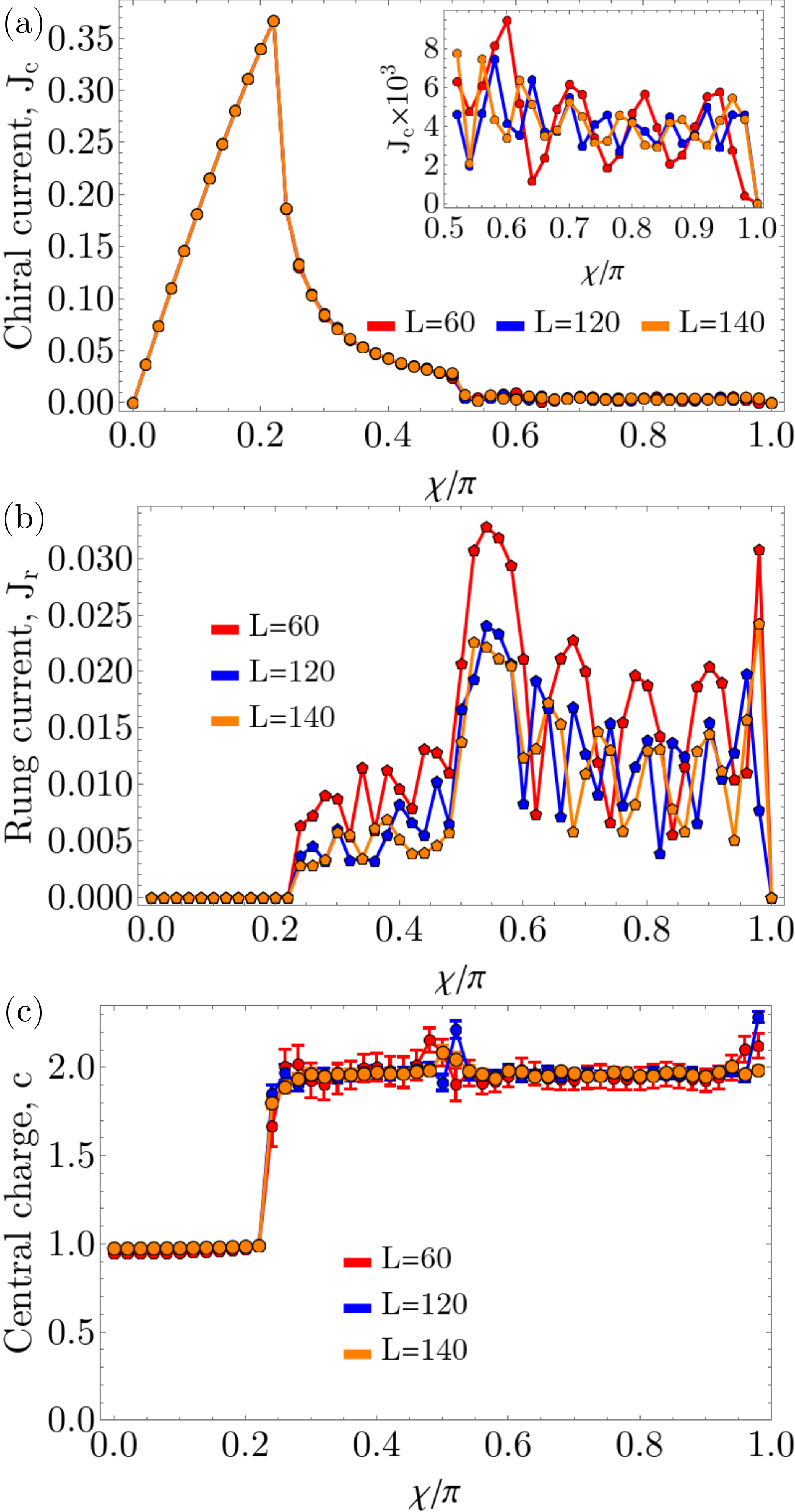}
\caption{Numerical ground state results for (a) the chiral current, $J_c$, (b) the average rung current, $J_r$, and (c) the central charge, $c$, as a function of the flux $\chi$ for hardcore bosons, $J_\|/J=2$, $\rho=0.5$, $L\in\{60,120,140\}$. We observe a transition between a Meissner superfluid and a vortex superfluid, at $\chi\approx0.23\pi$.
The maximal bond dimension used was $m=600$ for $L=60$, and $m=1200$ for $L=120$ and $L=140$.
 }
\label{fig:hardcore_jp2}
\end{figure}

\begin{figure}[!hbtp]
\centering
\includegraphics[width=.48\textwidth]{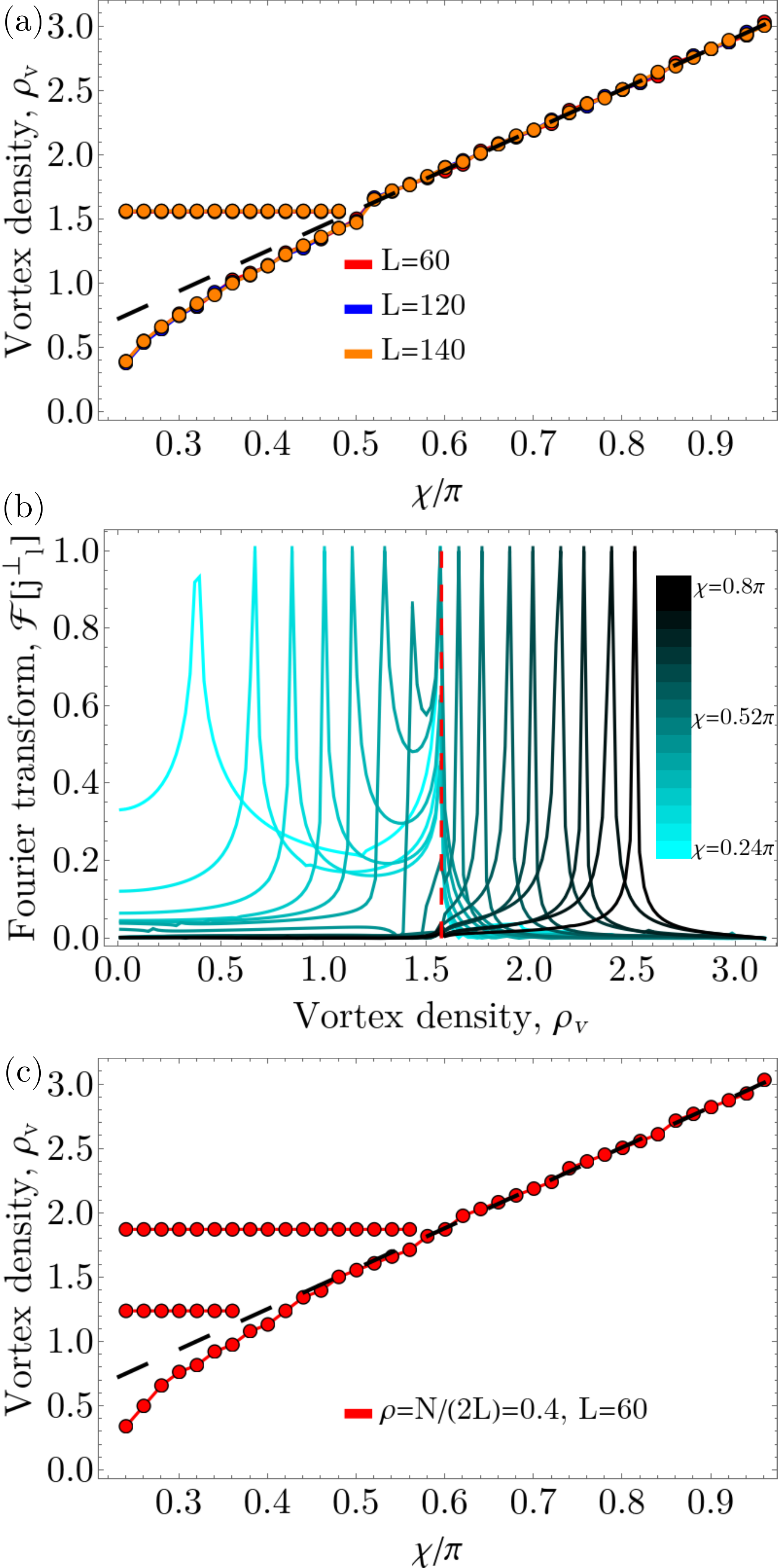}
\caption{Numerical ground state results for (a) the vortex density, $\rho_v$, at half filling as a function of the flux $\chi$ for hardcore bosons, $J_\|/J=2$, $\rho=0.5$, $L\in\{60,120,140\}$.
(b) The Fourier transform of the rung currents pattern $j^\perp_l$. The sharp peaks correspond to the extracted value of the vortex density. The dashed vertical line marks the value $\rho_v=\pi/2$. The parameters used are $J_\|/J=2$, $\rho=0.5$, $L=140$ and $0.24\pi\leq\chi\leq0.8\pi$.
(c) The vortex density, $\rho_v$, at filling $\rho=0.4$ as a function of the flux $\chi$ for hardcore bosons, $J_\|/J=2$, $L=60$.
The maximal bond dimension used was $m=600$ for $L=60$, and $m=1200$ for $L=120$ and $L=140$.
 }
\label{fig:hardcore_vortex}
\end{figure}

\begin{figure}[!hbtp]
\centering
\includegraphics[width=.48\textwidth]{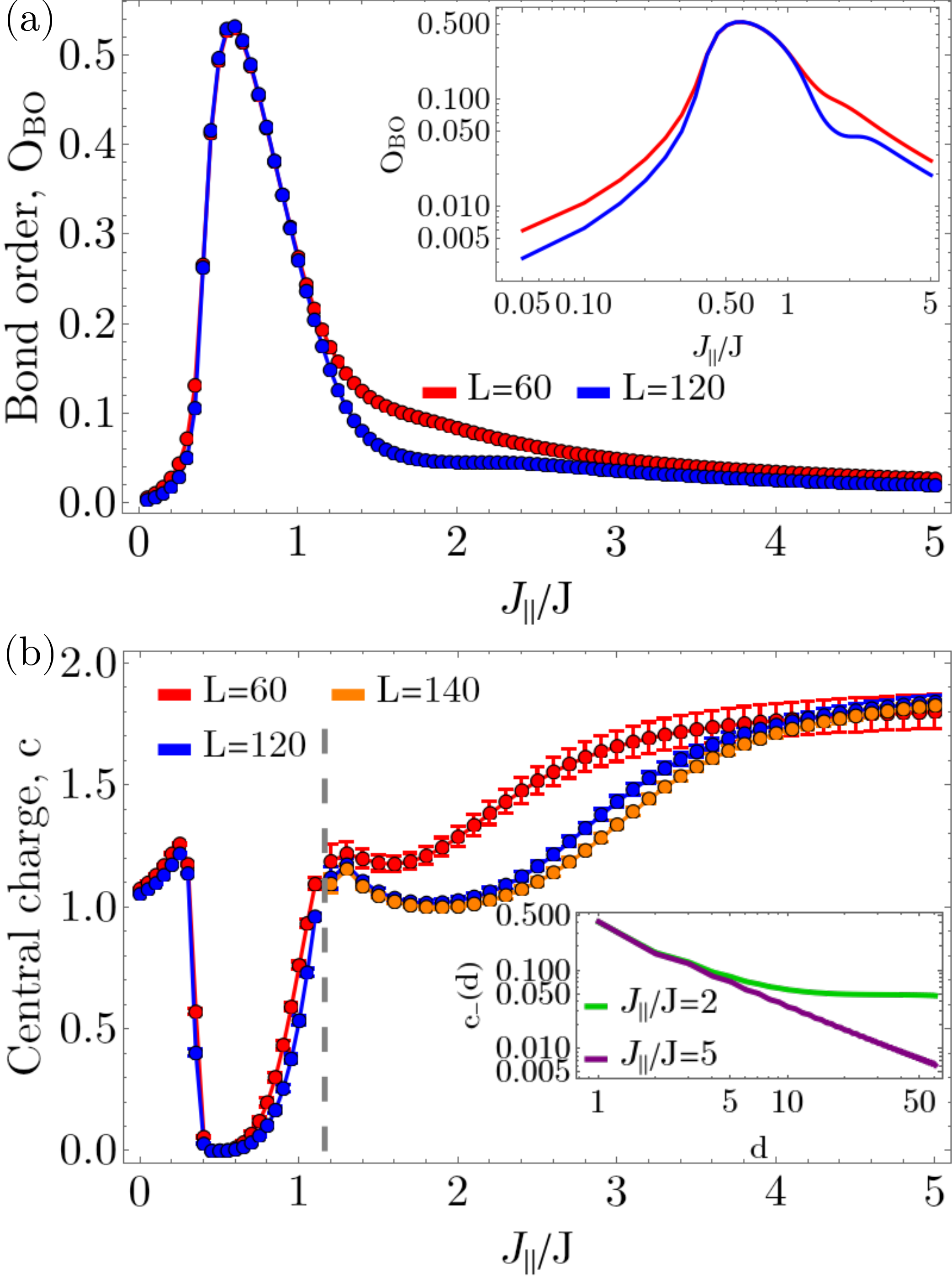}
\caption{Numerical ground state results for (a) the bond-order parameter, $O_\text{BO}$, and (b) the central charge, $c$, as a function of $J_\|/J$ for hardcore bosons, $\chi=\pi$, $\rho=0.5$, $L\in\{60,120,140\}$. 
For panel (b) above the vertical gray dashed line, $J_\|/J\gtrsim 1.2$, we employed finite currents for the first and final three sites, for breaking the $\mathbb{Z}_2$ symmetry in order to converge to the chiral superfluid phase, with amplitude $j_0/J=0.25$ (for more details see Appendix~\ref{app:dmrg_cblp}).
In the inset of panel (b) the rung-rung correlations $c_-(d)$ (\ref{eq:corr_pi}) are plotted as a function of distance in a log-log plot for $J_\|/J\in\{2,5\}$ and $L=120$.
We observe a transition from superfluid to bond-order insulating state at $J_\|/J\approx 0.3$, and from the insulating phase to the chiral superfluid at $J_\|/J\approx 1.2$. For large, $J_\|/J\gtrsim 4$, the system seems to be in a two mode superfluid, however the transition shows considerable finite size effects.
The maximal bond dimension used was $m=600$ for $L=60$, $m=1200$ for $L=120$, and $m=1500$ for $L=140$.
 }
\label{fig:hardcore_pi}
\end{figure}

In this section, we analyze the phase diagram in the case of hardcore bosons at half filling, $\rho=N/(2L)=0.5$. 

For $\chi<0.8\pi$, we obtain a phase diagram (see Fig.~\ref{fig:phasediag_hardcore}) similar for the finite interaction half-filling case, Sec.~\ref{sec:half_filling}, in which we observe a phase transition between a Meissner superfluid and a vortex superfluid.
The behavior of the chiral and rung currents are shown in Fig.~\ref{fig:hardcore_jp2}(a)-(b), for $J_\|/J=2$. In Fig.~\ref{fig:hardcore_jp2}(c) we see the jump from $c\approx1$ to $c\approx2$ in the central charge signaling the phase transition.
However, even though in the vortex phase we obtain finite values for the average rung current and chiral current for $\chi>0.5\pi$, the values are relatively small and show a strong dependence on the size of the system.

In contrast to the regime of finite on-site interaction, here in the vortex phase for $\chi<0.5\pi$, marked with gray in Fig.~\ref{fig:phasediag_hardcore}, we find two peaks in the Fourier transform of the space dependence of the rung currents [Fig.~\ref{fig:hardcore_vortex}(b)].
Above the value, $\chi=0.5\pi$, we find a single peak in the Fourier transform.
This implies that vortices of two different lengths coexist in the regime $\chi<0.5\pi$. We plot the corresponding vortex densities in Fig.~\ref{fig:hardcore_vortex}(a), where one branch corresponds to the expected value $\rho_v\approx\chi$, as discussed in Sec.~\ref{sec:twochain}. The second value of the vortex density seems to be related to the density of the atoms $\rho_v\approx0.5\pi=\rho \pi$.
We verify the dependence on the density by considering also the case of $\rho=0.4$ in Fig.~\ref{fig:hardcore_vortex}(c). Here we observe three different peaks, corresponding to $\rho_v\approx\chi$, $\rho_v\approx0.4\pi=\rho \pi$ up to $\chi=0.4\pi$, and $\rho_v\approx0.6\pi=(1-\rho) \pi$ up to $\chi=0.6\pi$.

In the hardcore limit for $\chi=\pi$, the model we consider has been analyzed in Ref.~\cite{MishraParamekanti2013}, furthermore, the model can be mapped to a frustrated spin chain, which has been studied in Refs.~\cite{Furukawa_2010_frustrated_spins, Furukawa_2012_frustrated_spins, Ueda_2020_frustrated_spins}
In the regime of small $J_\|/J$ a transition from a superfluid phase to a bond order insulator (BOI) has been pointed out.
The bond order insulator phase is characterized by a nonzero value of the bond order parameter
\begin{equation}
\label{eq:bondorder}
O_\text{BO}=\frac{1}{2L-1}\sum_j \left\langle b^\dagger_{j,1} b_{j,2}-b^\dagger_{j+1,1}b_{j,2}+\text{H.c.}\right\rangle.
\end{equation}
We observe the transition to the insulator phase at $J_\|/J\approx0.35$, signaled by the finite value of the bond order parameter and $c\approx 0$, as seen Fig.~\ref{fig:hardcore_pi}.

For $J_\|/J\gtrsim1.2$ a phase transition to the chiral superfluid phase is present, equivalent to the transition to the vector chiral phase observed in Refs.~\cite{Furukawa_2010_frustrated_spins, Furukawa_2012_frustrated_spins, Ueda_2020_frustrated_spins}.
In order to converge to the chiral superfluid phase we added a boundary term in the Hamiltonian which explicitly breaks the $\mathbb{Z}_2$ symmetry and favors the current pattern of the chiral phase (for details see Appendix~\ref{app:dmrg_cblp}).
Without the presence of such a term we obtain a state with vanishing currents as seen in Figs.~\ref{fig:hardcore_jp2}(a)-(b) for $J_\|/J=2$ and Fig.~\ref{fig:chiralphase} of Appendix~\ref{app:dmrg_cblp}.
Up to $J_\|/J\approx 3$ the numerically computed central charge is consistent with $c=1$ for the chiral superfluid phase [Fig.~\ref{fig:hardcore_pi}(b)]
and the rung-rung correlations $c_-(d)$ saturates at long distance, as seen in the inset of Fig.~\ref{fig:hardcore_pi}(b).
In contrast, for larger values of $J_\|/J$ it seems that the central charge has a value of $c\approx 2$ for the finite systems considered, even in the presence of the boundary terms. Thus, we obtain a state for which both the symmetric and antisymmetric sectors are gapless and identify as a two mode superfluid. This is supported also by the algebraic decay of the rung-rung correlations $c_-(d)$ on the length-scales considered [see inset of Fig.~\ref{fig:hardcore_pi}(b)].
However, due to important finite size effects seen in Fig.~\ref{fig:hardcore_pi}(b) and Fig.~\ref{fig:chiralphase} of Appendix~\ref{app:dmrg_cblp}, it is not clear from our finite size results if the two-mode superfluid is present in the thermodynamic limit, or the chiral superfluid will extend to arbitrary large $J_\|/J$.

\begin{figure}[!hbtp]
\centering
\includegraphics[width=.48\textwidth]{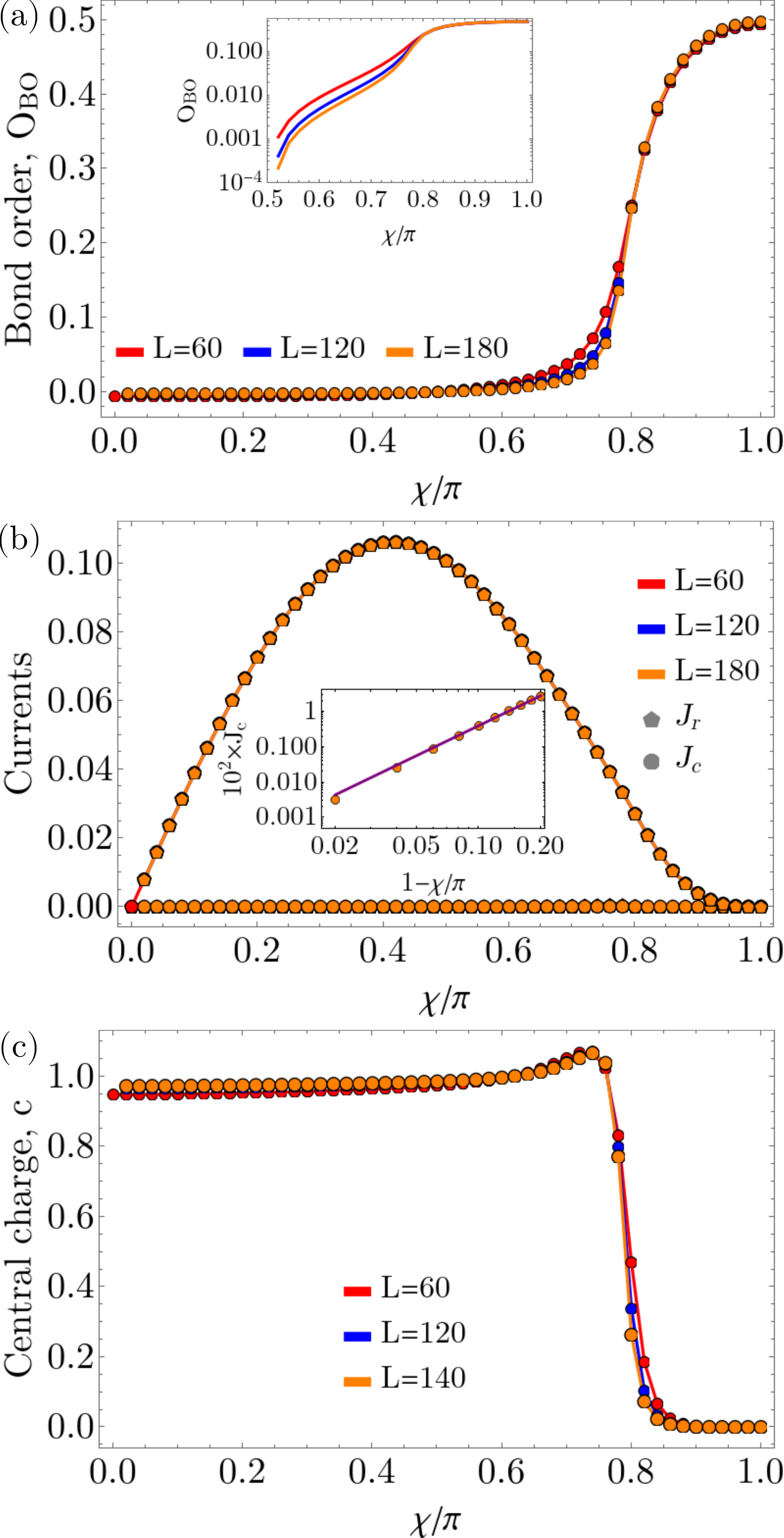}
\caption{Numerical ground state results for (a) the bond order parameter, $O_\text{BO}$,
(b) the chiral current, $J_c$, and average rung current, $J_r$, and (c) the central charge (c), at half filling as a function of the flux $\chi$ for hardcore bosons, $J_\|/J=0.5$, $\rho=0.5$, $L\in\{60,120,140\}$.
In the inset of panel (a) we plot the bond order parameter for $\chi>0.5\pi$ in a semi-log plot.
In the inset of panel (b) we show the algebraic scaling of the chiral current, $J_c$, as a function of $\chi_0=\pi-\chi$ in a log-log plot. The fitted power law, $\propto (\pi-\chi)^\alpha$, plotted in purple, corresponds to fitted exponent $\alpha=2.8\pm0.02$.
We can observe the transition between the Meissner superfluid and the chiral bond order insulato for $\chi\approx 0.8\pi$.
The maximal bond dimension used was $m=600$ for $L=60$, and $m=1400$ for $L=120$, and $m=1800$ for $L=140$.
 }
\label{fig:hardcore_bo}
\end{figure}

We observe that the bond order parameter remains finite also as we decrease the flux to $\chi<\pi$, as seen in Fig.~\ref{fig:hardcore_bo}(a) for $J_\|/J=0.5$. We can identify a transition between the bond order insulator and a Meissner superfluid for $\chi\approx 0.8\pi$, as the central charge has a jump from $c\approx 0$ to $c\approx 1$ lowering the value of the flux [Fig.~\ref{fig:hardcore_bo}(c)].
Interestingly, the bond order insulator exhibits finite values of the chiral current for $0.8\pi<\chi<\pi$ as seen in Fig.~\ref{fig:hardcore_bo}(b), with their pattern similar as in the Meissner phases, see Fig.~\ref{fig:currents_pattern}(e). This goes beyond the usual phenomenology of the bond order phase and, thus, we name this novel phase the chiral bond-order insulator.
We note that we observe very little finite size effects in the values of the chiral currents in increasing the size from $L=60$ to $L=180$ [see Fig.~\ref{fig:hardcore_bo}(b)].

For $J_\|/J=0.5$ and $\chi=\pi$ our model corresponds to the Majumdar-Gosh point \cite{MishraParamekanti2013}. For this point one can write exactly the ground state of the model as a product state
\begin{equation}
\label{eq:MG}
\ket{MG}=\prod_{j}\frac{1}{\sqrt{2}}\left(\ket{n_{j,1}=1, n_{j,2}=0}+\ket{n_{j,1}=0,n_{j,2}=1}\right).
\end{equation}
In the case of periodic boundary conditions a degenerate state exists, exhibiting the bond order to the rungs connecting the sites $(j+1,1)$ and $(j,2)$. Note that the ground state in this case is not a product of singlets like the usual Majumdar-Gosh state \cite{MishraParamekanti2013}.
Thus, in order to gain some insight into the chiral bond-order insulator we analyze the Hamiltonian in the case of $J_\|/J=0.5$ and  small $\chi_0=\pi-\chi$. The Hamiltonian reads
\begin{align} 
\label{eq:Hamiltonian_MG}
 H= & -\frac{J}{2} \sum_{j} \left(b^\dagger_{j}b_{j+1} +b^\dagger_{j}b_{j+1} -b^\dagger_{j}b_{j+2}+ \text{H.c}. \right)  \\
& +\frac{J}{2} \chi_0i\sum_j (-1)^j\left( b^\dagger_{j}b_{j+2} - \text{H.c}. \right), \nonumber
\end{align}
where we used the single chain representation (see Sec.~\ref{sec:singlechain}).
The first line of Hamiltonian (\ref{eq:Hamiltonian_MG}) corresponds to the Majumdar-Gosh points for which the state given in Eq.~(\ref{eq:MG}) is the ground state.
The second line resembles a current term, which can also be found in the expression of the chiral current. In this limit, the chiral current reads
\begin{align} 
\label{eq:Jc_MG}
 J_{c}= & \frac{J}{2}i \sum_j (-1)^j\left( b^\dagger_{j}b_{j+2} - \text{H.c}. \right)  \\
& +\frac{J}{2} \chi_0\sum_j \left( b^\dagger_{j}b_{j+2} + \text{H.c}. \right). \nonumber
\end{align}
In the numerical results we observe that the chiral current show an algebraic scaling with $\chi_0$ with an exponent of $\approx2.8$, see the inset of Fig.~\ref{fig:hardcore_bo}(b). This is consistent with the possibility that the term in the second line of  Eq.~(\ref{eq:Hamiltonian_MG}) would produce a higher order response and induce a finite value of the chiral current.

\begin{figure}[!hbtp]
\centering
\includegraphics[width=.48\textwidth]{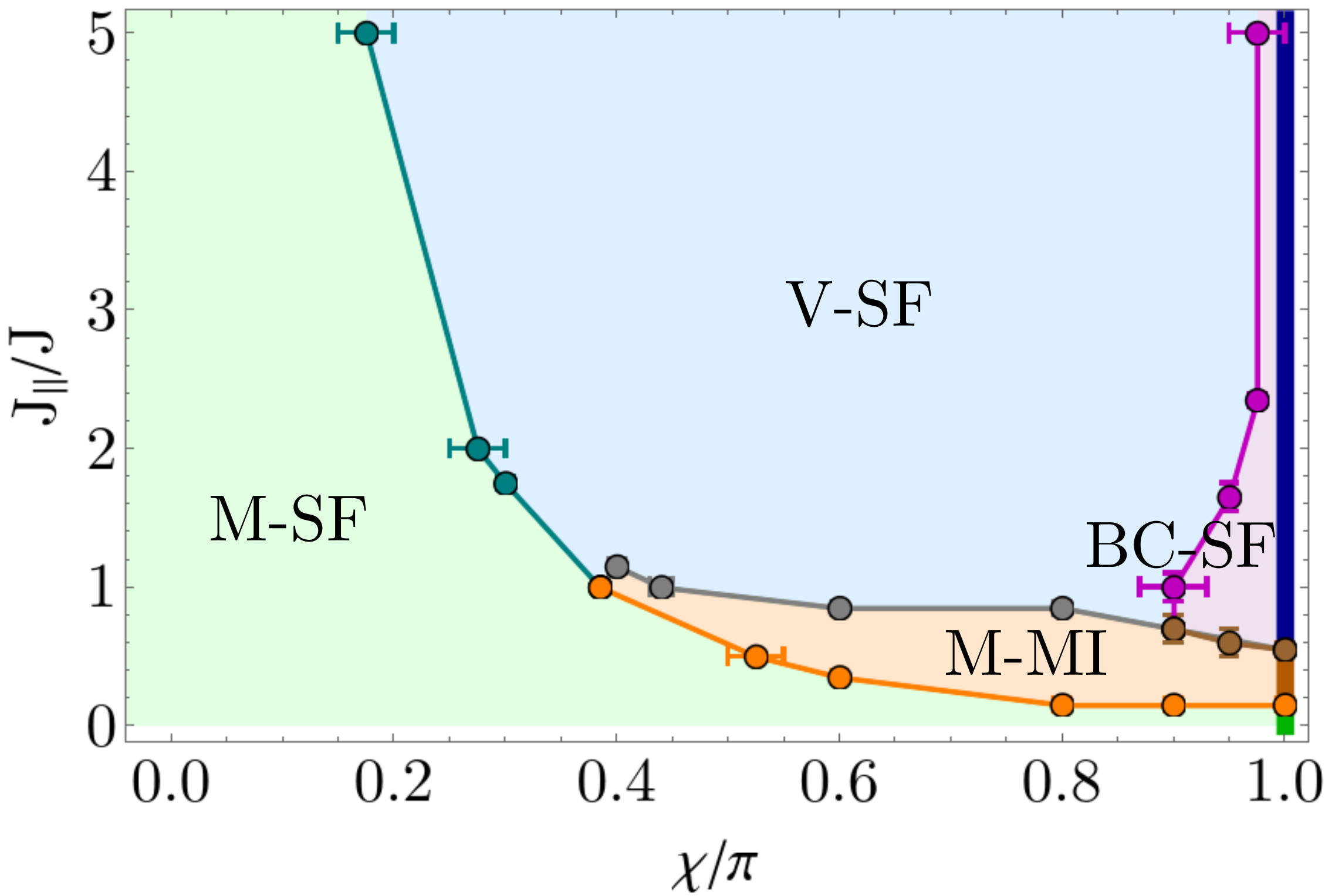}
\caption{Sketch of the phase diagram for unity filling, $\rho=1$, for $U/J=2.5$.
The identified phases (see text) are the Meissner Mott insulator (M-MI), the Meissner superfluid (M-SF), the vortex superfluid (V-SF) and biased chiral superfluid (BC-SF).
At $\chi=\pi$ (marked by a thick vertical line) we have phase transitions between a superfluid phase (green), a Mott insulator (brown) and a chiral superfluid (dark blue).
 }
\label{fig:phasediag_n1}
\end{figure}

\begin{figure}[!hbtp]
\centering
\includegraphics[width=.48\textwidth]{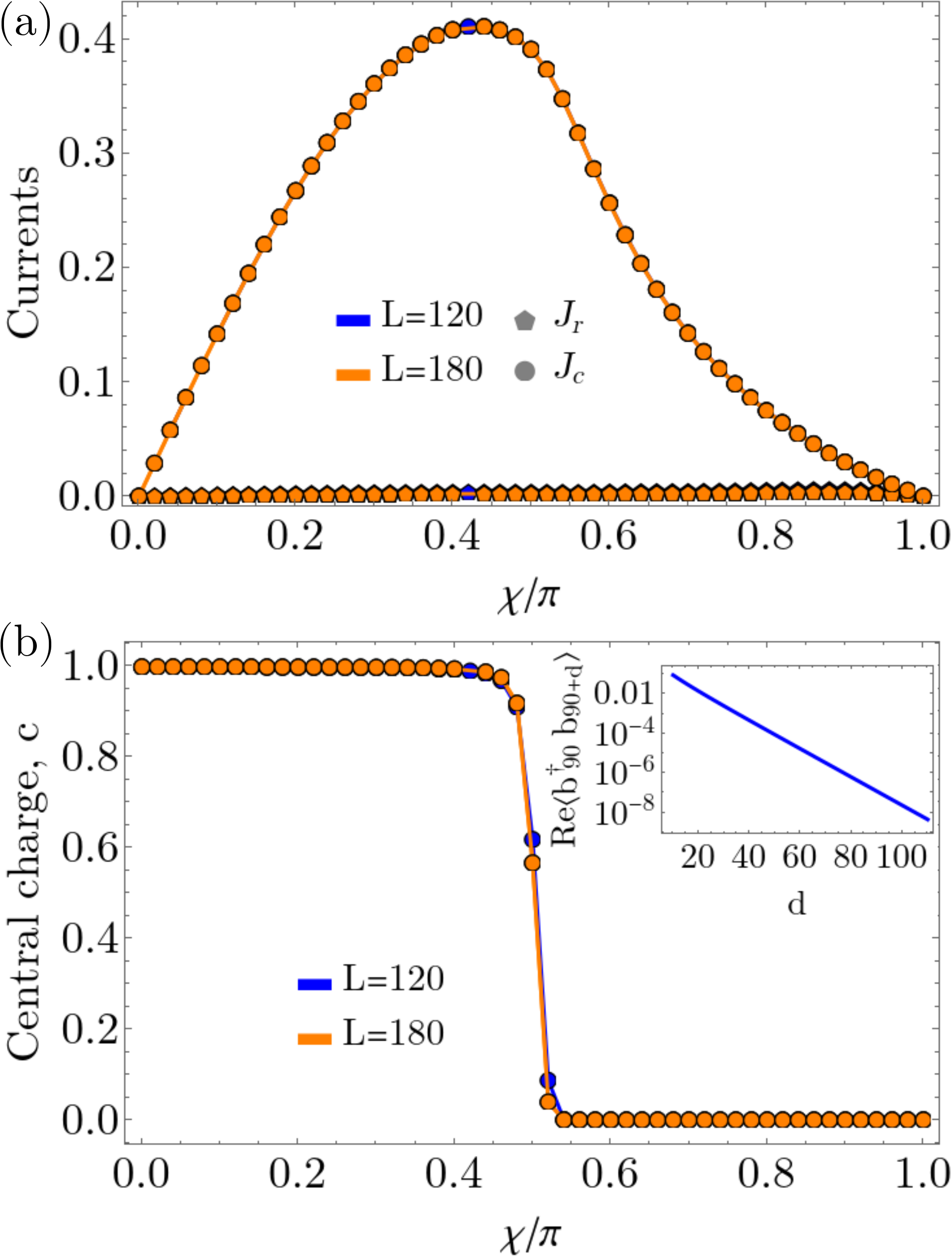}
\caption{Numerical ground state results for (a) the average rung current, $J_r$, and the chiral current, $J_c$, and (b) the central charge, $c$, as a function of the flux $\chi$ for $J_\|/J=0.5$, $U/J=2.5$, $\rho=1$, $L\in\{120,180\}$.
The inset of panel (b) contains the exponential decay with distance of the single particle correlations for $\chi= 0.56\pi$ and $L=180$.
We can identify a transition from the Meissner superfluid to the Meissner Mott insulator around $\chi\approx 0.5\pi$.
The maximal bond dimension used was $m=1200$ for $L=120$ and $m=1800$ for $L=180$.
 }
\label{fig:n1_msf_mmi}
\end{figure}

\section{Phase diagram at unity filling, $\rho=1$ \label{sec:unity_filling}}

\begin{figure}[!hbtp]
\centering
\includegraphics[width=.48\textwidth]{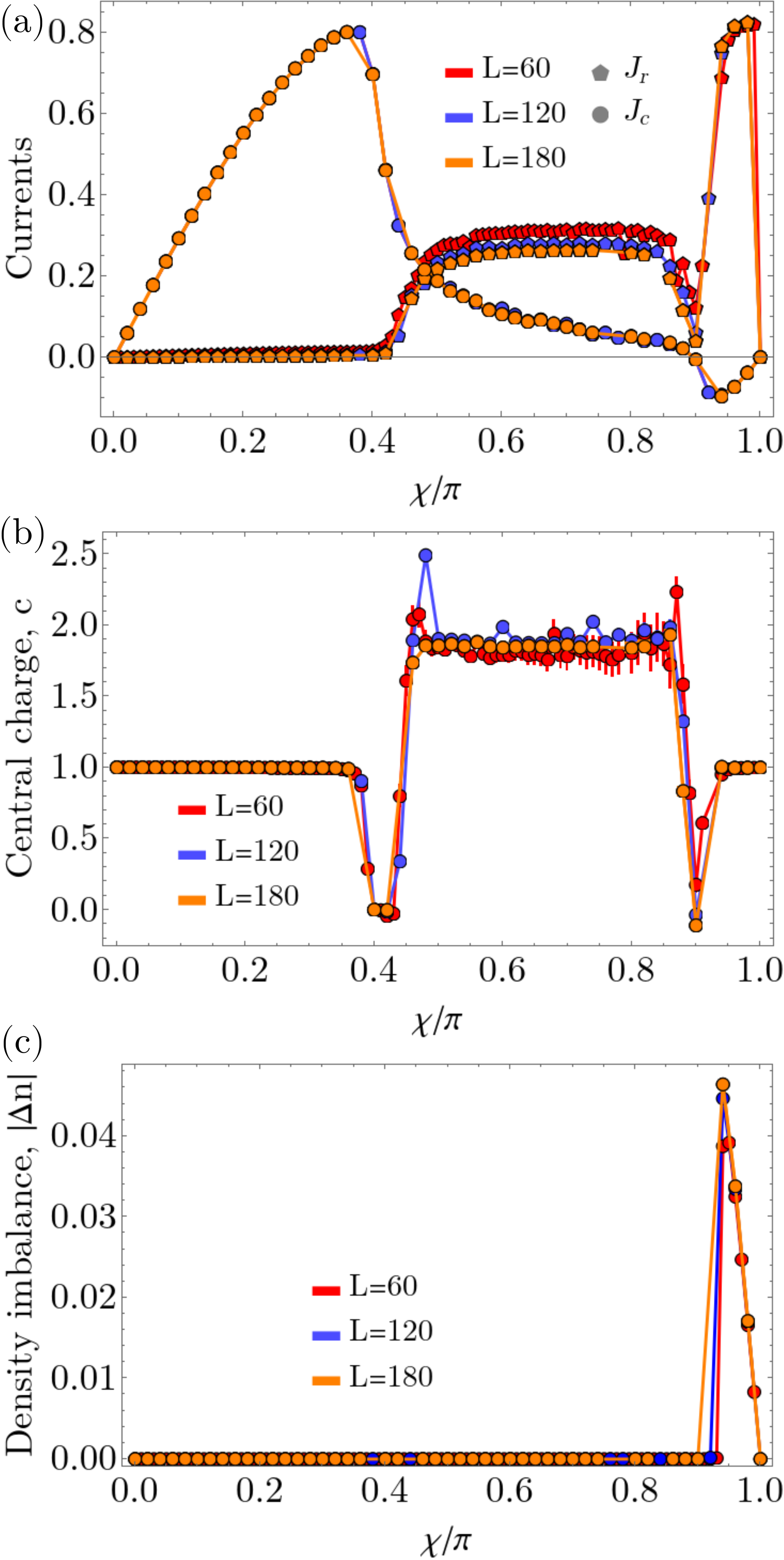}
\caption{Numerical ground state results for (a) the average rung current, $J_r$, and the chiral current, $J_c$, (b) the central charge, $c$, and (c) the absolute value of the density imbalance, $|\Delta n|$, as a function of the flux $\chi$ for $J_\|/J=1$, $U/J=2.5$, $\rho=1$, $L\in\{60,120,180\}$. We observe the following sequence of phases: Meissner superfluid, Meissner Mott-insulator, vortex superfluid and biased chiral superfluid. 
The maximal bond dimension used was $m=600$ for $L=60$, $m=1200$ for $L=120$ and $m=1800$ for $L=180$.
 }
\label{fig:n1_msf_mmi_vsf_bp}
\end{figure}

In this section, we analyze the phase diagram in the case in which we have a filling of one boson every site, $\rho=1$.
In this case the term $\cos\left[2p\phi(x)\right]$, with $p=1$, [second line of Eq.~(\ref{eq:Hamiltonian_chain_boson})], or its equivalent in Eq.~(\ref{eq:Hamiltonian_2chains_boson}), which stems from the commensurability effects of the interactions, may play an important role.
This causes the presence of the Meissner Mott-insulating phase (M-MI) in the phase diagram, Fig.~\ref{fig:phasediag_n1}.
Besides the presence of the insulating phase, we observe a very similar phase diagram compared to the half-filling case, Fig.~\ref{fig:phasediag_n05}.
Thus, in the following parts of this section we focus on the cuts through the phase diagram which include the Meissner Mott-insulator.

\begin{figure}[!hbtp]
\centering
\includegraphics[width=.48\textwidth]{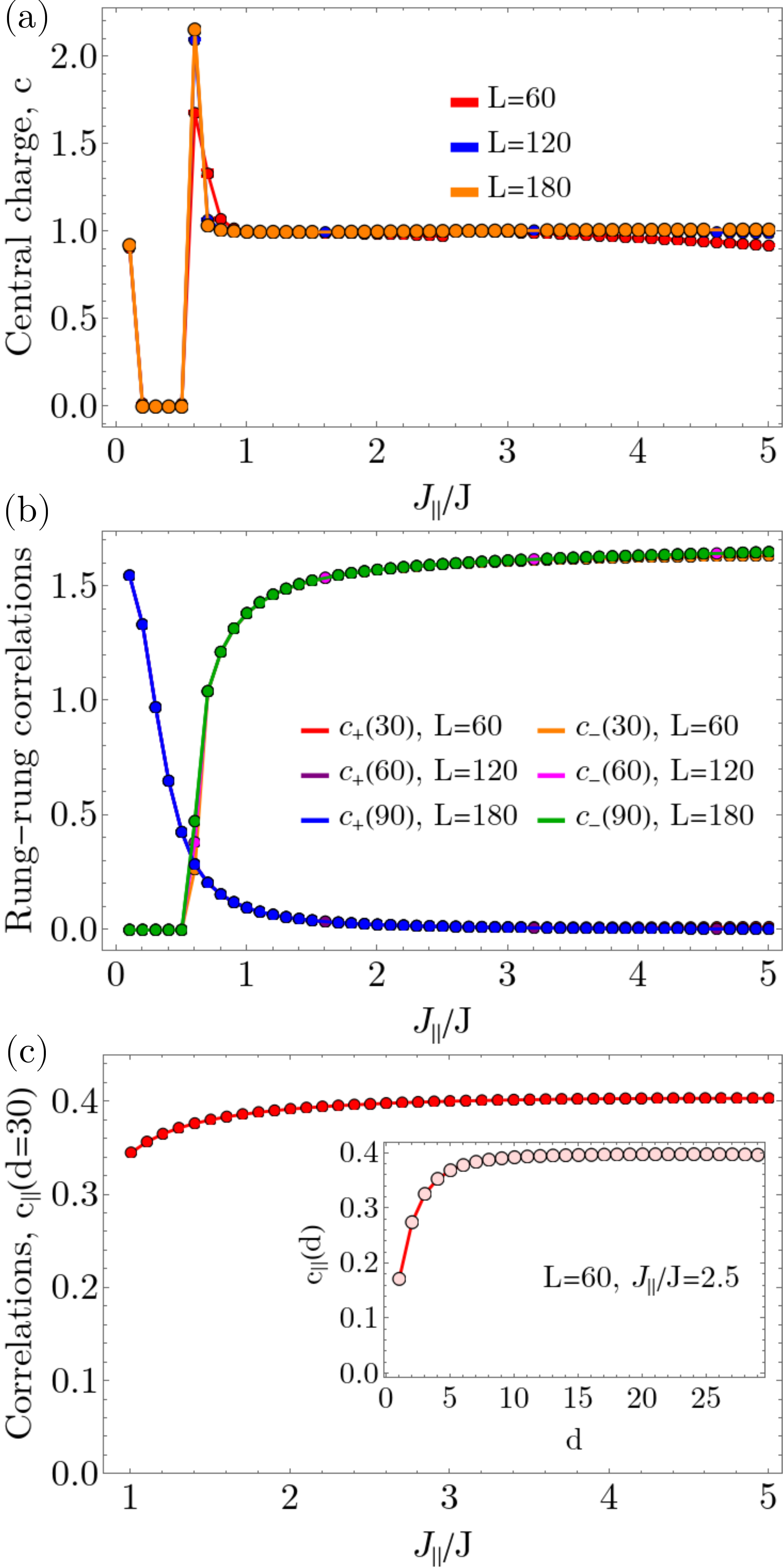}
\caption{Numerical ground state results for (a) the central charge, (b) the saturation value of the rung-rung correlations $c_\pm$ (\ref{eq:corr_pi}), (c) the saturation value of the current-current correlations on the legs $c_\|$ (\ref{eq:corr_pi_2}), as a function of $J_\|$, for $\chi=\pi$, $U/J=2.5$, $\rho=1$, $L\in\{60,120,180\}$.
In the inset of panel (c) we show the distance dependence of $c_\|(d)$ at $J_\|/J=2.5$.
We observe a transition from superfluid to Mott insulator states at $J_\|/J\approx 0.15$, and from the insulating phase to the chiral superfluid at $J_\|/J\approx 0.6$.
The maximal bond dimension used was $m=500$ for $L=60$, $m=1200$ for $L=120$ and  $m=1500$ for $L=180$.
 }
\label{fig:n1_pi}
\end{figure}

In Fig.~\ref{fig:n1_msf_mmi}, we show the transition between the Meissner superfluid and the Meissner Mott-insulator for $J_\|/J=0.5$. In this regime the preferred description is that of a single chain (\ref{eq:Hamiltonian_chain})-(\ref{eq:Hamiltonian_chain_boson}).
In both phases we have a strong chiral current on the legs of the ladder and no currents on the rungs, Fig.~\ref{fig:n1_msf_mmi}(a). With the pattern of currents corresponding to the one depicted in Fig.~\ref{fig:currents_pattern}.
However, around $\chi\approx 0.5\pi$ we see that the central charge [Fig.~\ref{fig:n1_msf_mmi}(b)] goes from a value of $c=1$ to $c=0$, signaling the transition to the insulating phase.
As the Meissner Mott-insulator is a fully gapped phase one expects that $c=0$.
Furthermore, the single particle correlations decay exponentially in this phase, as seen in the inset of Fig.~\ref{fig:n1_msf_mmi}(b) for $\chi= 0.56\pi$.

Similarly to the half-filling case, Sec.~\ref{sec:half_filling}, by increasing $J_\|/J$ we can also obtain the vortex superfluid phase, see Fig.~\ref{fig:phasediag_n1}. In Fig.~\ref{fig:n1_msf_mmi_vsf_bp}, for $J_\|/J=1$, we observe that for $\chi\lesssim 0.45$ we have a strong chiral current and no currents on the rungs, signaling the Meissner superfluid phase up to $\chi\lesssim 0.39\pi$ and the Meissner Mott-insulator for $0.39\pi\lesssim\chi\lesssim 0.45\pi$, as the central charge changes from $c\approx1$ to $c\approx0$ around $\chi\approx0.39\pi$.
At $\chi\approx0.45\pi$ a transition to the vortex superfluid occurs, as the central charge becomes $c\approx2$ and the rungs currents are finite.
For large values of the flux, $\chi\gtrsim 0.9\pi$ we enter the biased chiral superfluid phase, which shows a finite density imbalance, Fig.~\ref{fig:n1_msf_mmi_vsf_bp}(c). The biased phase is present up to large values of $J_\|/J$ for $\chi$ close to $\pi$, Fig.~\ref{fig:phasediag_n1}.

At large $J_\|/J$ for $\chi=\pi$ we obtain the chiral superfluid phase, for which $c=1$, the correlations $c_-$ and $c_\|$ saturate at a finite value at large distances indicating the strong currents present on both the rungs and legs of the ladder, see Fig.~\ref{fig:n1_pi}.
As we decrease $J_\|/J$ a transition to a Mott-insulator phase is present at  $J_\|/J\approx0.6$, as seen in the vanishing value of the central charge $c=0$ [Fig.~\ref{fig:n1_pi}(a)]. For a value of $J_\|/J=0.1$ the central charge is close to one, signaling the superfluid phase.

\section{Discussion \label{sec:discussions}}

In this section, we discuss the obtained phase diagrams in comparison with the phases seen in square ladders, with an emphasis on the novel phases occurring in the triangular ladder. To recall the phase diagrams see Fig.~\ref{fig:phasediag_n05} for $\rho=0.5$, Fig.~\ref{fig:phasediag_hardcore} for $\rho=0.5$ and hardcore bosons, and Fig.~\ref{fig:phasediag_n1} for $\rho=1$.

At small values of the flux, $\chi\lesssim0.6\pi$, the behavior is essentially similar to the one of square ladders and exhibits phase transition between Meissner and vortex states \cite{OrignacGiamarchi2001}.
We obtain both the superfluid and Mott insulator states with Meissner character. However, whether vortex Mott-insulating phases can be observed in the triangular ladder, as in the square ladder \cite{GreschnerVekua2016,PetrescuLeHur2013} remains an open question.
For hardcore bosons, we identify a new effect in the vortex phase, namely the presence of a second frequency peak in the Fourier transform of the pattern of rung currents, which seems to be commensurate with the bosonic density. The explanation of such a harmonic is unclear at the moment and will clearly deserve further studies. 

Contrary to what happens for small flux, at large values of $\chi$, the frustration induced by the triangular nature of the hopping becomes more prominent and novel phases appear.
One phase without an equivalent on the square ladder is the chiral bond order insulator, which we obtained in the hardcore limit. This phase is different from other states exhibiting bond ordering due to its finite chiral current flowing on the legs of the triangular ladder.
This bond ordered phase does not stem from a band insulator limit for small $J_\|/J$, as for example the Meissner Mott insulator present at half filling for a square ladder \cite{PiraudSchollwoeck2015}. This can be inferred from the fact that at small $J_\|/J$ we have a transition to a Meissner superfluid, or superfluid for $\chi=\pi$ (see Fig.~\ref{fig:phasediag_hardcore}).

At finite values of the on-site interaction we find another novel phase: the biased chiral superfluid, a phase breaking the discrete $\mathbb{Z}_2$ symmetry of the ladder.
Even though in the weakly interacting limit this phase can be understood as the condensation of the bosons in a single minimum of the double minima potential, similarly to the biased ladder phase appearing in the square ladder
\cite{WeiMueller2014, UchinoTokuno2015, GreschnerVekua2016}, the nature of its currents is very different. The biased phase of the square ladder exhibits Meissner-like currents, but the biased chiral superfluid is closely related to the chiral superfluid present at $\chi=\pi$. 
Thus, due to the frustration for $\chi$ close to $\pi$, the symmetric and antisymmetric sectors are coupled, such that a gapped antisymmetric sector implies a finite value of the expectation value of the gradient of the symmetric field, $\langle\partial_x\theta_s\rangle$. This mechanism induces strong currents flowing on opposite direction on the rungs and legs of the triangular ladder [see Fig.~\ref{fig:currents_pattern} (c)-(d)].

One interesting direction left open by our study is the investigation of the behavior of the phases with increasing interaction strength. In particular, there is the question at $\rho=0.5$ on how to connect the phase diagram for $U/J=2.5$, Fig.~\ref{fig:phasediag_n05}, and the phase diagram for $U/J\to\infty$, Fig.~\ref{fig:phasediag_hardcore}.
Our preliminary numerical data shows that the chiral bond order insulator extends to finite interactions, as low as $U/J\approx10$ for $J_\|/J=0.5$ and $\chi=0.95\pi$, where a phase transition to the biased chiral superfluid might be present.
As the extent of the biased chiral superfluid seems to diminish as we increase $U$, it would interesting to see if the phase will be suppressed at a critical value of the interaction, or if it survives up to the hardcore $U\to\infty$ limit. 
Furthermore, in the case of $\rho=1$, as we expect the Meissner Mott insulator to cover a larger region of the phase diagram as we increase $U$, the question arises if any other phases remain stable at very large interaction strengths.

Ultracold atoms in optical lattices provide an experimental platform for the study of such low-dimensional systems in the presence of an artificial gauge field.
The flux has been implemented with time-dependent modulations \cite{Struck_2012_shakenlattice}, laser-assisted Raman hopping \cite{AidelsburgerBloch2011, MiyakeKetterle2013, Tai2017_2body_flux}, or synthetic dimensions \cite{ManciniFallani2015, Genkina_2019_Hofstadterribbons, Chalopin_2020_Hall2D, Zhou_2022_Hallexperiment, Roell_2022_ErbiumHall}.
Combining these techniques with triangular optical lattices \cite{Becker_2010_triangularlattice,StruckSengstock2011} offers the possibility of the experimental realization of the model we studied.

In order to distinguish the different ground state phases one could perform different measurements.
From insitu measurements one could access the local densities and currents \cite{Buser_2032_currentssnapshot}, and the momentum distribution can be obtained via time-of-flight measurements.
In the case of square ladders, in Ref.~\cite{AtalaBloch2014}, Atala et al. (2014) used a measurement scheme involving the projection onto double wells to measure the chiral current.
The biased chiral superfluid phase seems to be robust for densities away from half-filling, thus, it should be possible to observe it also in the case of a parabolic trapping potential. However, for the bond ordered states having one particle every two sites is important.

\section{Conclusions \label{sec:conclusions}}

In this work, we investigated the phase diagram of interacting bosonic atoms confined to a two-leg triangular ladder in an artificial gauge field.
We showed the existence of Meissner phases both in the superfluid (M-SF) and in the Mott insulator (M-MI) regimes and of incommensurate vortex superfluid phases (V-SF), which are similar to the states obtained in square ladders with a flux.
However, we have seen that at large values of the flux the frustration effects of the triangular geometry play a crucial role and several novel phases are realized.
For finite on-site interaction a biased chiral superfluid (BC-SF) phase which breaks the $\mathbb{Z}_2$ symmetry of the ladder and exhibits an imbalanced density on the two legs of the ladder is present. It has a similar current pattern with the chiral superfluid phase (C-SF) which is obtained as the $\chi\to\pi$ limit of the biased phase.
We study the transition from the chiral superfluid to the superfluid phase by performing an interpolation between the $\chi=\pi$ triangular ladder and the square ladder.
In the case of hardcore bosons, we show the presence of a chiral bond order insulator (C-BOI) phase, which corresponds to a finite value of the bond order parameter and exhibits Meissner-like currents.

Our work paves the way for studies of the triangular ladder with an artificial gauge field in non-equilibrium settings. This can be envisioned either in the context of the Hall effect \cite{Greschner_2019_HallEffect,Buser_2021_HallEffect}, or by the coupling to an external environment \cite{Guo_2016_transport_ladder,KollathBrennecke2016, HalatiKollath2017,HalatiKollath2019,Xing_2020_transport_ladder}.

\section*{ACKNOWLEDGMENTS}
We thank M.~Aidelsburger, L.~Pizzino and L.~Tarruell for stimulating discussions. 
We are grateful to S.~Furukawa for pointing out the presence of the chiral superfluid phase in the hardcore case and for enlightening discussions on that point.
This work was supported by the Swiss National Science Foundation under Division II grant 200020-188687

\section*{APPENDIX}

\setcounter{section}{0}
\renewcommand{\thesection}{\Alph{section}}
\renewcommand{\theequation}{A.\arabic{equation}}
\setcounter{equation}{0}

\section{Non-interacting limit\label{app:noninteracting}}

We diagonalize the kinetic part of the Hamiltonian
\begin{align} 
\label{eq:app_Hamiltonian_kin}
 H_\text{kin}=& H_\perp+H_\parallel,
\end{align}
by performing the Fourier transforms on the two legs
\begin{align} 
\label{eq:app_Fourier}
b_m(k) = \frac{1}{\sqrt{L}} \sum_j e^{i k j} b_{j,m}.
\end{align}
The Hamiltonian in momentum space reads
\begin{align} 
\label{eq:app_Hamiltonian_mom}
 H_\text{kin}=&\sum_k \Big\{ -2 J_\| \cos(k+\chi) b_1^\dagger (k) b_1 (k) \\
 &-2 J_\| \cos(k-\chi) b_2^\dagger (k) b_2 (k) \nonumber \\
 &- J\left[\left(1+e^{ik}\right) b_1^\dagger (k) b_2 (k)+\left(1+e^{-ik}\right)b_2^\dagger (k) b_1 (k)\right]\Big\}. \nonumber
\end{align}
The eigenvalues of this quadratic Hamiltonian are
\begin{align} 
\label{eq:app_eigenvalues}
E_\pm (\chi)=&-2 J_\| \cos(k)\cos(\chi) \\
&\pm\sqrt{2J^2\left[1+\cos(k)\right]+4J_\|^2\sin^2(k)\sin^2(\chi)}. \nonumber
\end{align}
The energy bands are plotted in Fig.~\ref{fig:bands}.

If we compare these bands with the ones obtained for the square lattice of the double of the flux
\begin{align} 
\label{eq:app_eigenvalues_square}
E_\pm^\text{square} (2\chi)=&-2 J_\| \cos(k)\cos(\chi) \\
&\pm\sqrt{J^2+4J_\|^2\sin^2(k)\sin^2(\chi)}, \nonumber
\end{align}
we observe that in the limit of large $J_\|$ they become very similar. Thus, one can expect in this regime analogous behaviors in the two setups.

\renewcommand{\theequation}{B.\arabic{equation}}
\setcounter{equation}{0}

\section{Luttinger parameter in the single chain limit \label{app:K_1chain}}

\begin{figure}[!hbtp]
\centering
\includegraphics[width=.48\textwidth]{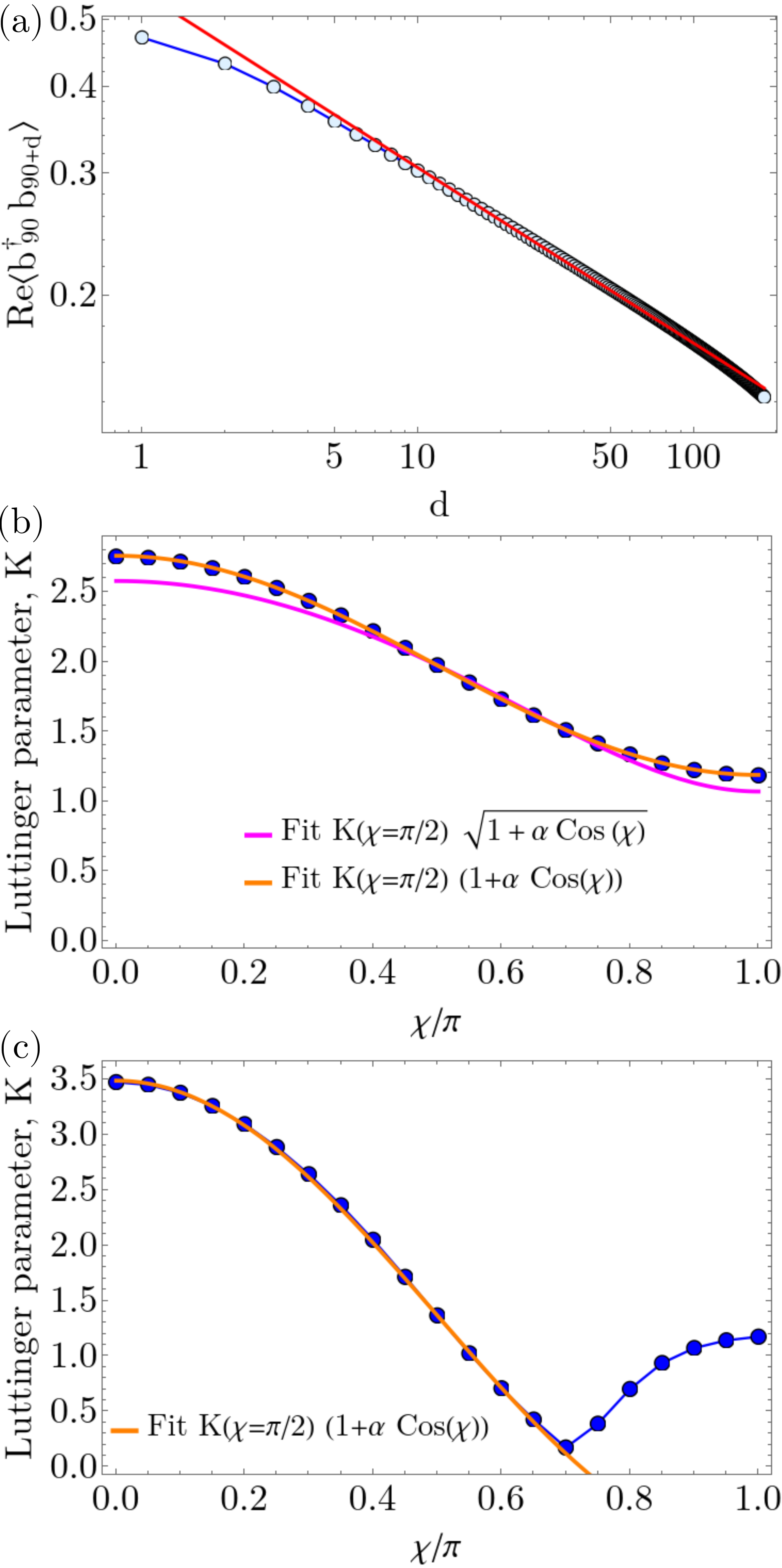}
\caption{(a) The decay of the single particle correlations, $\Re\left\langle b^\dagger_{90} b_{90+d}\right\rangle$, in the Meissner superfluid, for $\chi=0.5\pi$, $J_\|/J=0.2$, $U/J=2.5$, $N/2L=0.5$, $L=180$. The algebraic fit follows Eq.~(\ref{eq:K_bb}) and corresponds to the fitted value of $K=1.973\pm0.005$.
(b)-(c) The dependence of the Luttinger parameter $K$ as a function of the flux $\chi$, for (b) $J_\|/J=0.2$, (c)$J_\|/J=0.5$, $U/J=2.5$, $L=60$ and $N/2L=0.5$. The solid curves correspond to fitting $K(\chi=\pi/2)+\alpha\cos(\chi)$ (\ref{eq:K_chain2}) (orange) and $K(\chi=\pi/2) \sqrt{1+\alpha\cos(\chi)}$ (\ref{eq:K_chain2_num}) (magenta).
 }
\label{fig:luttK}
\end{figure}

For a Bose-Hubbard chain (\ref{eq:Hamiltonian_chain})-(\ref{eq:Hamiltonian_chain_boson}) one can numerically compute the Luttinger parameter of the superfluid phase by looking at the algebraic decay of the single particle correlations \cite{Giamarchibook}
\begin{align} 
\label{eq:K_bb}
\langle b^\dagger(x) b(0)\rangle\propto x^{-1/(2K)},
\end{align}
where $x$ is the lattice position in the continuum limit. In Fig.~\ref{fig:luttK}(a) we show the decay of the single particle correlations for parameters corresponding the Meissner superfluid phase and its algebraic fit.

In Eq.~(\ref{eq:Hamiltonian_chain_boson}) we saw that the effective Luttinger parameter of the model depends on the value of the flux 
\begin{align} 
\label{eq:K_chain}
 u_\text{eff} K _\text{eff} &= uK+t_\| \cos(\chi) \\
\frac{u_\text{eff}}{K_\text{eff}} &= \frac{u}{K}, \nonumber
\end{align}
where $t_\|\equiv16\pi \rho J_\|$, which leads to the following dependence on the flux
\begin{align} 
\label{eq:K_chain2}
 K _\text{eff} = K \sqrt{1+\frac{t_\|}{uK}\cos(\chi)}.
\end{align}
However, the numerical results seem to deviate from this dependence for the values of the flux close to the maxima of $|\cos(\chi)|$, Fig.~\ref{fig:luttK}. 
Furthermore, we observe that the numerical data follows 
\begin{align} 
\label{eq:K_chain2_num}
K_\text{eff} = K+\alpha\cos(\chi),
\end{align}
as seen in Fig.~\ref{fig:luttK}(b). 
We note that the discrepancy between the two dependencies becomes smaller if we consider smaller values of $t_\|$.

In the case the long-range hopping in the chain is strong enough, $t_\|>uK$, for large values of the flux the prefactor $uK+t_\| \cos(\chi)$ can become zero or negative signaling an instability. We observe the approach towards zero also in the dependence of the numerically extracted Luttinger parameter. This marks the phase transition between the Meissner superfluid to the biased phase.

\renewcommand{\theequation}{C.\arabic{equation}}
\setcounter{equation}{0}

\section{Mean-field approach for the $\mathbb{Z}_2$ symmetry broken phase \label{app:mf_cblp}}

Starting from the Hamiltonian derived in Sec.~\ref{sec:2chain_pi}
\begin{align} 
\label{eq:Hamiltonian_mf}
 H&=H_0+ H_\perp \\
 H_0&= \int\frac{dx}{2\pi}\left[uK\partial_x\theta_s(x)^2+\frac{u}{K}\partial_x\phi_s(x)^2\right] \nonumber \\
 &+ \int\frac{dx}{2\pi}\left[uK\partial_x\theta_a(x)^2+\frac{u}{K}\partial_x\phi_a(x)^2\right] \nonumber \\
  H_\perp&=-t_\perp \int dx \sin\left(\sqrt{2}\theta_a\right) \partial_x \theta_s, \nonumber 
\end{align}
where $t_\perp=\sqrt{2}\rho J$, we want to investigate the effect of the $H_\perp$ term which couples the symmetric and antisymmetric sectors. For this, we follow the self-consistent mean-field procedure described in Ref.~\cite{Nersesyan_1998_zigzagspinladder} for frustrated spin ladders.

We perform the mean-field decoupling of $H_\perp$ to obtain the following mean-field Hamiltonian density
\begin{align} 
\label{eq:Hamiltonian_mf_2}
 \mathcal{H}_\text{MF}&=\mathcal{H}_0+ k \partial_x \theta_s -\mu \Lambda \sin\left(\sqrt{2}\theta_a\right),
\end{align}
where we have assumed that the ground state of the system has a non-zero current $\partial_x \theta_s$. The self-consistency conditions are given by 
\begin{align} 
\label{eq:Hamiltonian_mf_3}
k &= -t_\perp \left\langle \sin\left(\sqrt{2}\theta_a\right) \right\rangle, \\
\mu \Lambda &= t_\perp  \left\langle \partial_x \theta_s \right\rangle. \nonumber
\end{align}
One can observe that $\mathcal{H}_\text{MF}$ decomposes in two independent parts
\begin{align} 
\label{eq:Hamiltonian_mf_4}
 \mathcal{H}_\text{s}&=uK\partial_x\theta_s(x)^2+\frac{u}{K}\partial_x\phi_s(x)^2+\partial_x \theta_s, \\
  \mathcal{H}_\text{a}&=uK\partial_x\theta_a(x)^2+\frac{u}{K}\partial_x\phi_a(x)^2-\mu \Lambda \sin\left(\sqrt{2}\theta_a\right). \nonumber
\end{align}
In the symmetric sector we can arrive at a quadratic Hamiltonian by performing the following redefinition of the field, $\theta_s(x)\to\theta_s(x)-\frac{k}{2uK} x$. The average value of  $\theta_s(x)$ is, thus, given by 
\begin{align} 
\label{eq:Hamiltonian_mf_5}
 \left\langle \partial_x \theta_s(x)\right\rangle=-\frac{k}{2uK}.
\end{align}
The Hamiltonian for the antisymmetric sector is a sine-Gordon model for the field $\theta_a$, this can be solved exactly and the value of the mass is given by \cite{Giamarchibook}
\begin{align} 
\label{eq:Hamiltonian_mf_6}
\left\langle \sin\left(\sqrt{2}\theta_a\right) \right\rangle=c \left(\frac{\mu}{uK\Lambda}\right)^\frac{1}{4K-1},
\end{align}
where $c$ is a constant, which can be calculated \cite{Lukyanov_1997_exactsineGordon}.
Inserting the expectation values from Eqs.~(\ref{eq:Hamiltonian_mf_5})-(\ref{eq:Hamiltonian_mf_6}) in Eq.~(\ref{eq:Hamiltonian_mf_3}) we obtain the solutions
\begin{align} 
\label{eq:Hamiltonian_mf_7}
\mu&=\Tilde{c} t_\perp ^\frac{4K-1}{2K-1} (uK\Lambda)^{-\frac{2K}{2K-1}} \\
k&= -2 \Tilde{c} t_\perp ^\frac{2K}{2K-1} (uK\Lambda)^{-\frac{1}{2K-1}}, \nonumber
\end{align}
where $\Tilde{c}$ is a constant. In the regime in which the solutions are valid the field $\theta_a$ is fixed to the minima of the sine-Gordon potential, $\langle \theta_a \rangle =\frac{\pi}{2\sqrt{2}}$ and the mass in the antisymmetric sector scales as $\left\langle \sin\left(\sqrt{2}\theta_a\right) \right\rangle\propto t_\perp ^\frac{1}{2K-1}$.

\begin{figure}[!hbtp]
\centering
\includegraphics[width=.48\textwidth]{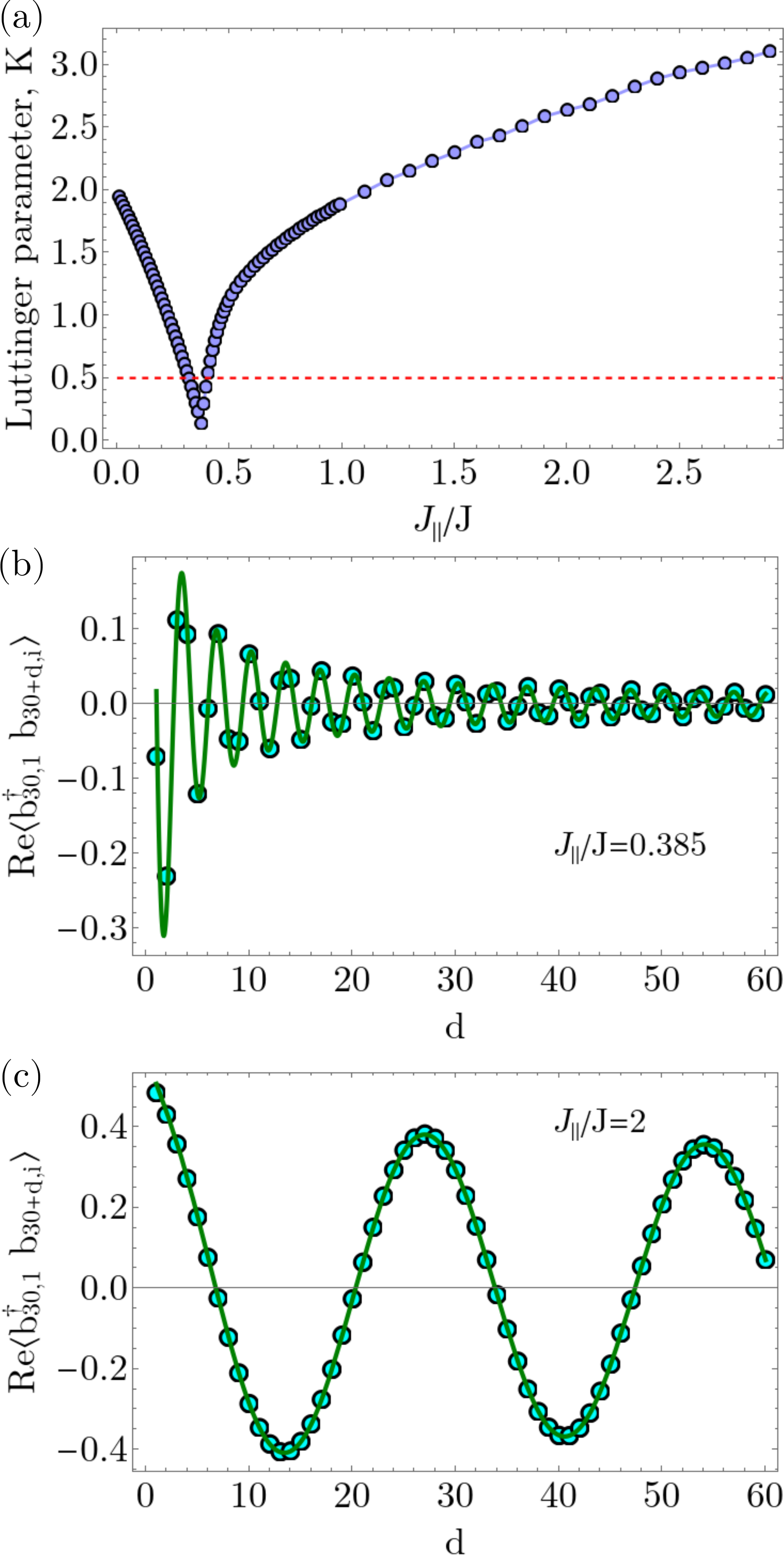}
\caption{(a) The dependence of the Luttinger parameter $K$ on $J_\|/J$ for $\chi=\pi$, $U/J=2.5$, $N/2L=0.5$, $L=120$. We extract $K$ from the decay of the single particle correlations, using Eq.~(\ref{eq:K_bb}) in the superfluid phase, $J_\|/J\lesssim 0.37$ and Eq.~(\ref{eq:mf_corr_1}) in the chiral superfluid phase, $J_\|/J\gtrsim 0.37$. The dashed red line corresponds to $K=0.5$.
(b)-(c) The decay of the single particle correlations, $\Re\left\langle b^\dagger_{30,1} b_{30+d,i}\right\rangle$, in the chiral superfluid for (b) $J_\|/J=0.385$ and (c) $J_\|/J=2$. The solid line corresponds to a fit with the functional form of Eq.~(\ref{eq:mf_corr_1}), the fitted Luttinger parameters are (b) $K=0.294 \pm 0.001$ and (c) $K=2.64 \pm 0.01$.
 }
\label{fig:luttK_pi}
\end{figure}

The Hamiltonian from which we started (\ref{eq:Hamiltonian_mf}) has a $\mathbb{Z}_2$ symmetry as the Hamiltonian remains invariant under the following change of signs $\partial_x \theta_s\to -\partial_x \theta_s$ and $\sin\left(\sqrt{2}\theta_a\right)\to -\sin\left(\sqrt{2}\theta_a\right)$, which changes the expectation value of the field $\langle \theta_a \rangle =\frac{\pi}{2\sqrt{2}}\to \langle \theta_a \rangle =\frac{3\pi}{2\sqrt{2}}$. 
This implies that changing the signs in the mean-field solution still satisfies the self-consistency conditions (\ref{eq:Hamiltonian_mf_3}).
Thus, we obtain that in the regime in which the self-consistency conditions have solutions the ground state is given by two degenerate states which break the $\mathbb{Z}_2$ symmetry.

The solutions of the self-consistency conditions (\ref{eq:Hamiltonian_mf_7}) can be obtain when $K>0.5$. However, in the numerical results, around the phase transition between the superfluid and the chiral superfluid, $0.23\lesssim J_\|/J\lesssim 0.47$, we obtain a $K<0.5$ also in the chiral superfluid phase [see Fig.~\ref{fig:luttK_pi}(a)]. We compute numerically $K$ from the decay of the single particle correlations [see Fig.~\ref{fig:luttK_pi}(b)-(c)].
We attribute this discrepancy to the fact that we numerically observe $K>0.5$ for small values of $J_\|$, for which the approach of considering the system as two coupled chains might not be valid.

Using the mean-field results we can compute the scaling of the single particle correlations along one of the legs of the ladder
\begin{align} 
\label{eq:mf_corr_1}
\left\langle b^\dagger_{0,1}b_{j,1} \right\rangle&\propto \left\langle e^{i\left(\theta_1(x)-\theta_1(0)\right)} \right\rangle\\
&\propto x^{-\frac{1}{4K}}e^{-i \frac{k}{2\sqrt{2}uK} x}, \nonumber
\end{align}
where we can see that the correlations exhibit incommensurate oscillations. We plotted the numerical behavior of the correlations in Fig.~\ref{fig:luttK_pi}(b)-(c) for two values of $J_\|/J$ in the chiral superfluid phase.
We observe that the scaling of Eq.~(\ref{eq:mf_corr_1}) agrees very well with the numerical results.

In the rest of this section we extend the mean-field approach to the Hamiltonian (\ref{eq:Hamiltonian_pi_square})-(\ref{eq:Hamiltonian_perp_square}) we used to investigate the interpolation to the square ladder in Sec.~\ref{sec:2chain_square}.
\begin{align} 
\label{eq:Hamiltonian_mf_sq}
 H=&H_0+ H_\perp \\
 H_0=& \int\frac{dx}{2\pi}\left[uK\partial_x\theta_s(x)^2+\frac{u}{K}\partial_x\phi_s(x)^2\right] \nonumber \\
 &+ \int\frac{dx}{2\pi}\left[uK\partial_x\theta_a(x)^2+\frac{u}{K}\partial_x\phi_a(x)^2\right] \nonumber \\
  H_\perp=&-t_\perp \int dx \sin\left(\sqrt{2}\theta_a\right) \partial_x \theta_s\nonumber \\
  &-\tilde{t} \int dx  \cos\left(\sqrt{2}\theta_a\right), \nonumber 
\end{align}
where $\tilde{t}=2\rho t$. The mean-field decoupling is performed in the same manner as previously and the self-consistency conditions are the same as in Eq.~(\ref{eq:Hamiltonian_mf_3}). The difference arises in the potential present in the antisymmetric Hamiltonian density
\begin{align} 
\label{eq:Hamiltonian_mf_sq_1}
  \mathcal{H}_\text{a}=&uK\partial_x\theta_a(x)^2+\frac{u}{K}\partial_x\phi_a(x)^2\\
  &-\mu \Lambda \sin\left(\sqrt{2}\theta_a\right)-\tilde{t} \cos\left(\sqrt{2}\theta_a\right)\nonumber \\
  =&uK\partial_x\theta_a(x)^2+\frac{u}{K}\partial_x\phi_a(x)^2\nonumber \\
  &-\sqrt{(\mu \Lambda)^2+\tilde{t}^2} \sin\left(\sqrt{2}\theta_a+\psi\right), \nonumber
\end{align}
with the phase $\psi=\arctan\left(\frac{\mu \Lambda}{\tilde{t}}\right)$. The mass in the sine-Gordon model is given by \cite{Giamarchibook}
\begin{align} 
\label{eq:Hamiltonian_mf_sq_2}
\left\langle \sin\left(\sqrt{2}\theta_a+\psi\right) \right\rangle=c \left(\frac{\sqrt{(\mu \Lambda)^2+\tilde{t}^2}}{uK\Lambda}\right)^\frac{1}{4K-1}.
\end{align}
This leads to the self-consistent solution
\begin{align} 
\label{eq:Hamiltonian_mf_sq_3}
(\mu\Lambda)^2&=\Tilde{c} t_\perp ^\frac{8K-2}{2K-1}-\tilde{t}^2.
\end{align}
The mean-field approach allows us to compute the scaling with the coupling constants of the following expectation values
\begin{align} 
\label{eq:Hamiltonian_mf_sq_corr}
&\left\langle\sin\left(\sqrt{2}\theta_a\right)\right\rangle\propto t_\perp^{-2} \sqrt{\Tilde{c}_1 t_\perp ^\frac{8K-2}{2K-1}-\tilde{t}^2}, \\
&\left\langle\sin\left[\sqrt{2}\theta_a(x\to\infty)\right]\sin\left[\sqrt{2}\theta_a(0)\right]\right\rangle\propto \nonumber\\
&\qquad\qquad\qquad~\propto t_\perp ^{-4} \left(\Tilde{c}_1 t_\perp ^\frac{8K-2}{2K-1}-\tilde{t}^2\right), \nonumber\\
&\left\langle\cos\left(\sqrt{2}\theta_a\right)\right\rangle\propto \tilde{t} t_\perp^{-2}, \nonumber\\
&\left\langle\cos\left(\sqrt{2}\theta_a(x\to\infty)\right)\cos\left(\sqrt{2}\theta_a(0)\right)\right\rangle\propto \tilde{t}^2 t_\perp ^{-4}.\nonumber
\end{align}

\renewcommand{\theequation}{D.\arabic{equation}}
\setcounter{equation}{0}

\section{Numerical results in the $\mathbb{Z}_2$ symmetry broken phases \label{app:dmrg_cblp}}

\begin{figure}[!hbtp]
\centering
\includegraphics[width=.48\textwidth]{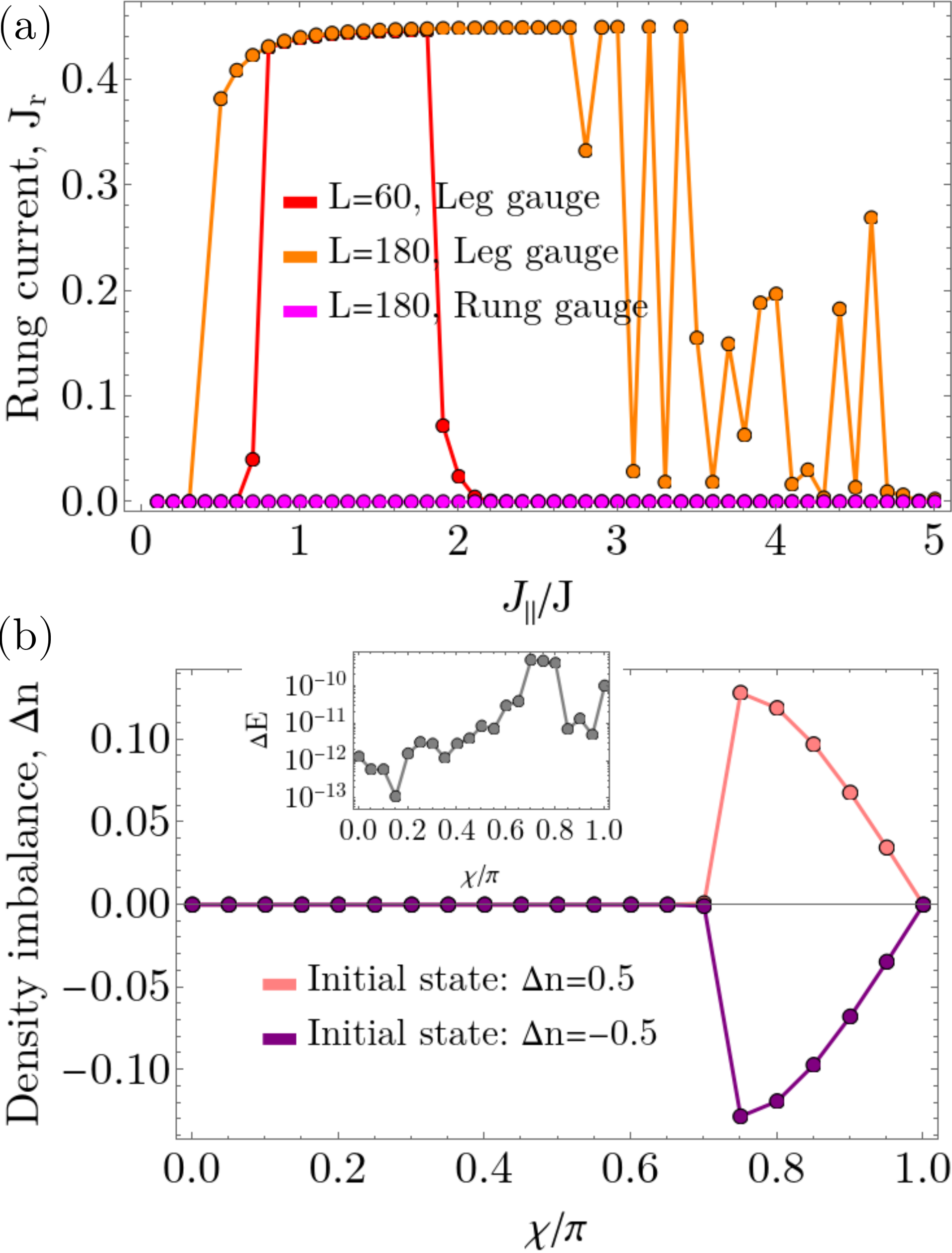}
\caption{(a) The average rung current, $J_r$, as a function of $J_\|/J$ for two system sizes, $L=60$ and $L=180$, the numerical simulations were done with the Hamiltonian either in the leg gauge or in the rung gauge (see main text, Sec.~\ref{sec:2chain_pi}). The parameter used are $\chi=\pi$, $U/J=2.5$, $N/2L=0.5$.
(b) The density imbalance, $\Delta n$, as a function of the flux $\chi$ for two different initial states used in the numerical DMRG simulations.
The inset contains the energy difference, $\Delta E$, between the two numerical results.
The parameter used are $J_\|/J=0.5$, $U/J=2.5$, $N/2L=0.5$ and $L=60$.
 }
\label{fig:biasedphase}
\end{figure}

\begin{figure}[!hbtp]
\centering
\includegraphics[width=.48\textwidth]{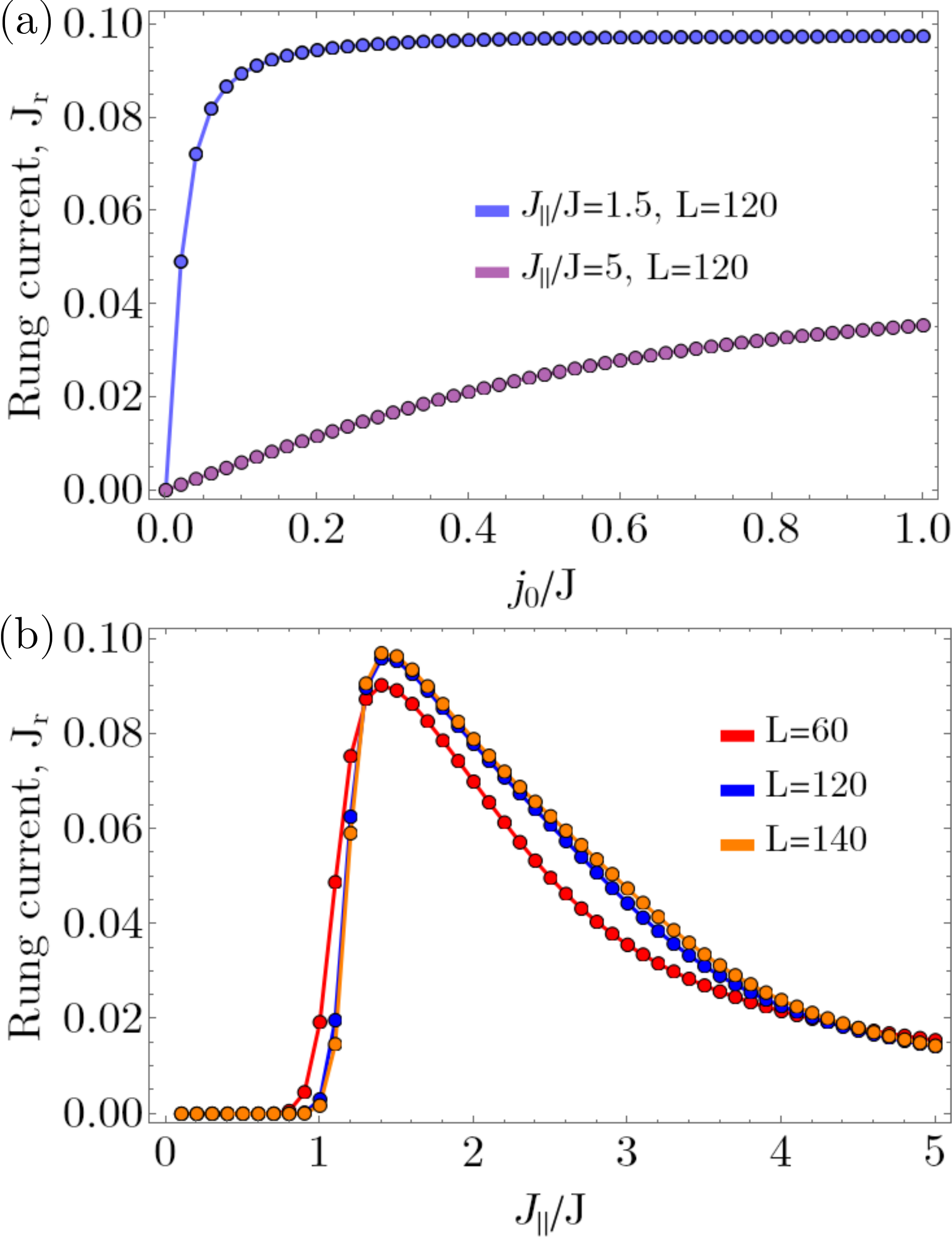}
\caption{The average rung current, $J_r$, computed in the middle half of the system for hardcore bosons, (a) as a function of the strength of the boundary currents $j_0/J$ (\ref{eq:edge}) for $J_\|/J\in\{1.5,5\}$;
(b) as a function of $J_\|/J$ for $L\in\{60,120,140\}$, with finite boundary currents $j_0/J=0.25$.
The parameters used are $\chi=\pi$, $U/J\to\infty$ and $\rho=0.5$.
The maximal bond dimension used was $m=600$ for $L=60$, $m=1200$ for $L=120$, and $m=1500$ for $L=140$.
 }
\label{fig:chiralphase}
\end{figure}

In this section, we briefly discuss the challenges of the numerical convergence in the cases in which we have two degenerate ground states due to the breaking of the $\mathbb{Z}_2$ symmetry. The DMRG algorithm can in principle converge to any state of the two-dimensional ground state manifold, this can increase the difficulty of the identification of the nature of the phases.
In this work we identified two such phases, the chiral superfluid for $\chi=\pi$ (see Sec.~\ref{sec:2chain_pi}) and the biased chiral superfluid for small values of $\chi_0=\pi-\chi$. (see Sec.~\ref{sec:singlechain} and Sec.~\ref{sec:close_to_pi}).

The difficulties of interpreting the numerical results for the average rung current are apparent when the ground state search algorithm uses the Hamiltonian in the leg gauge (\ref{eq:Hamiltonian_pi}) for $\chi=\pi$ and $L=60$ as seen in Fig.~\ref{fig:biasedphase}(a).
For these parameters we identified in Sec.~\ref{sec:2chain_pi} a phase transition between the superfluid and chiral superfluid phases at $J_\|/J\approx 0.3$. However, the rung current [red points in Fig.~\ref{fig:biasedphase}(a)] seems to indicate two transitions at $J_\|/J\approx 0.7$ and $J_\|/J\approx 2$, which we do not expect based on our analytical consideration.
This behavior can be explained by the possibility that for $0.7\lesssim J_\|/J\lesssim 2$ we numerically converge to one of the chiral states with strong currents described in Appendix~\ref{app:mf_cblp} and for the other values of $J_\|/J$ to an equal superposition of the two states with opposite current patterns.
Furthermore, if we analyze a large system, $L=180$, in the same gauge we observe that at large $J_\|/J$ the values obtained for the rung current seem to strongly depend on the value of $J_\|/J$, [orange points in Fig.~\ref{fig:biasedphase}(a)], which would imply that for each point we obtain different weights of the superposition.

On the other hand, in the rung gauge (\ref{eq:Hamiltonian_pi_transformed}), we converge to a state with zero currents for all parameters considered [magenta points in Fig.~\ref{fig:biasedphase}(a)]. This is consistent with an equal superposition of the two states with opposite current patterns. 
We note that we used the rung gauge for the results at $\chi=\pi$ presented in the main text.
Thus, in order to be able to identify the chiral superfluid phase in a confident manner, based on the insights obtain from the mean-field approach of Appendix~\ref{app:mf_cblp}, we computed in Sec.~\ref{sec:2chain_pi} the rung-rung correlations (\ref{eq:corr_pi}) and current-current correlations along the legs (\ref{eq:corr_pi_2}) which have the same value for any state in the ground state manifold.

Furthermore, we analyze the effect of the initial states in the ground state search algorithm. We observe in Fig.~\ref{fig:biasedphase}(b) that in the region in which we identified the biased chiral superfluid in Sec.~\ref{sec:singlechain} the value of the density imbalance depends on the initial state.
Here we used an initial state a product state in which all atoms were either on the first leg, $\Delta n=0.5$, or on the second leg, $\Delta n=-0.5$.
Furthermore, the two numerically obtained states have the same energy as seen in the inset of Fig.~\ref{fig:biasedphase}(b). Our implementation of the DMRG algorithm guarantees the convergence of the ground state energy up to $10^{-8}$.
We note that the results presented in Fig.~\ref{fig:biasedphase}(b) do not imply that we converged to the state with a maximal (or minimal) value of the density imbalance. However, it does show that we obtained two distinct states in the ground state manifold which supports the conclusion that we break the $\mathbb{Z}_2$ symmetry in this regime.

An approach to facilitate the convergence of the DMRG algorithm to one of the $\mathbb{Z}_2$ symmetry broken states is to add a term in the Hamiltonian that breaks the symmetry.
Afterwards, one extrapolates the results in the limit of strength of the symmetry breaking term going to $0$. For example, in order to identify the chiral superfluid phase for hardcore bosons at $\chi=\pi$ in Sec.~\ref{sec:hardcore} [see Fig.~\ref{fig:hardcore_pi}(b)], we employed the following term in the simulations (written in the single chain representation, Eq.~\ref{eq:Hamiltonian_chain})
\begin{align} 
\label{eq:edge}
 H_b= -i j_0 \left(\sum_{j=1}^3+\sum_{j=L-3}^{L-1}\right)\left(b^\dagger_{j}b_{j+1}-b^\dagger_{j+1}b_{j}\right),
\end{align}
which favors the rung current pattern realized in the chiral superfluid and biased chiral superfluid phases at the boundaries of the system.
In Fig.~\ref{fig:chiralphase} we analyze the behavior of the currents around the chiral superfluid phase when using boundary currents in the simulations.
The average rung current in the bulk of the system, computed in the middle half, as a function of $j_0/J$ is shown in Fig.~\ref{fig:chiralphase}(a). We can observe that for $J_\|/J=1.5$ the rung current rapidly saturates with increasing $j_0/J$, and, thus, we can be confident in the presence of the chiral superfluid phase by extrapolating in the limit of $j_0/J \to 0$. However, for $J_\|/J=5$ the extrapolation seems to rather indicate a state without currents.
In Fig.~\ref{fig:chiralphase}(b) we show the average rung current in the bulk of the ladder as a function of $J_\|/J$ for a fixed strength of $j_0/J=2.5$. 
We observe that for $1.2\lesssim J_\|/J\lesssim 4$, the rung current increases with increasing the system size, we attribute this behavior to the presence of the chiral superfluid phase (see Sec.~\ref{sec:hardcore}).
It is not easy, for the considered system sizes and value of $j_0/J$, to distinguish in Fig.~\ref{fig:chiralphase}(b) if a phase transition to a two-mode superfluid occurs above $J_\|/J\gtrsim 4$. For this a careful analysis of the extrapolation and system size dependence is needed.

\end{document}